\documentclass[aps,prd,10pt,tightenlines,notitlepage,nofootinbib,superscriptaddress]{revtex4-1}

\usepackage{amsfonts,amssymb,amsthm,bbm,amsmath,wasysym}
\usepackage{color,psfrag}
\usepackage[dvips]{graphicx}

\newcommand{\C}{{\mathbb C}}
\newcommand{\N}{{\mathbb N}}
\newcommand{\R}{{\mathbb R}}

\newcommand{\cA}{{\mathcal A}}
\newcommand{\cE}{{\mathcal E}}
\newcommand{\cF}{{\mathcal F}}
\newcommand{\cG}{{\mathcal G}}

\newcommand{\cV}{{\mathcal V}}
\newcommand{\cR}{{\mathcal R}}
\newcommand{\cP}{{\mathcal P}}
\newcommand{\cH}{{\mathcal H}}
\newcommand{\cM}{{\mathcal M}}

\newcommand{\cC}{{\mathcal C}}
\newcommand{\cX}{{\mathcal X}}

\newcommand{\SU}{\mathrm{SU}}

\newcommand{\SL}{\mathrm{SL}}
\newcommand{\SO}{\mathrm{SO}}
\newcommand{\U}{\mathrm{U}}

\newcommand{\be}{\begin{equation}}
\newcommand{\ee}{\end{equation}}
\newcommand{\beq}{\begin{eqnarray}}
\newcommand{\eeq}{\end{eqnarray}}
\newcommand{\bea}{\begin{eqnarray}}
\newcommand{\eea}{\end{eqnarray}}
\newcommand{\nn}{\nonumber}

\newcommand{\mat} [2] {\left ( \begin{array}{#1}#2\end{array} \right ) }

\newcommand{\su}{{\mathfrak{su}}}

\renewcommand{\u}{{\mathfrak u}}

\newcommand{\la}{\langle}
\newcommand{\ra}{\rangle}

\newcommand{\tr}{{\mathrm Tr}}
\newcommand{\f}{\frac}

\def\nn{\nonumber}
\def\pp{\partial}
\def\arr{\rightarrow}

\def\eps{\epsilon}

\newcommand{\id}{\mathbb{I}}

\def\tz{\tilde{z}}

\def\tZ{\tilde{Z}}
\def\tX{\tilde{X}}
\def\tcX{\tilde{\cX}}

\def\tF{\tilde{F}}
\def\tg{\tilde{g}}
\def\bz{\bar{z}}

\def\bcX{\bar{\cX}}
\def\bF{\overline{F}}
\def\bE{\overline{E}}
\def\vV{\vec{V}}
\def\vW{\vec{W}}
\def\vcC{\vec{\cC}}
\def\vJ{\vec{J}}
\def\vsigma{\vec{\sigma}}
\def\hE{\hat{E}}
\def\hF{\hat{F}}

\def\te{\tilde{e}}
\def\tv{\tilde{v}}

\begin{document}

\title{Deformation Operators of Spin Networks and Coarse-Graining}

\author{{\bf Etera R. Livine}}\email{etera.livine@ens-lyon.fr}
\affiliation{Laboratoire de Physique, ENS Lyon, CNRS-UMR 5672, 46 All\'ee d'Italie, Lyon 69007, France}
\affiliation{Perimeter Institute, 31 Caroline St N, Waterloo ON, Canada N2L 2Y5}

\date{\today}

\begin{abstract}

In the context of loop quantum gravity, quantum states of geometry are mathematically  defined as spin networks living on graphs embedded in the canonical space-like hypersurface. In the effort to study the renormalisation flow of loop gravity, a necessary step is to understand the coarse-graining of these states in order to describe their relevant structure at various scales.
Using the spinor network formalism to describe the phase space of loop gravity on a given graph, we focus on a bounded (connected) region of the graph and coarse-grain it to a single vertex using a gauge-fixing procedure.
We discuss the ambiguities in the gauge-fixing procedure and its consequences for coarse-graining spin(or) networks.
This allows to define the boundary deformations of that region in a gauge-invariant fashion and to identify the area preserving deformations as $\U(N)$ transformations similarly to the already well-studied case of a single intertwiner. The novelty is that the closure constraint is now relaxed and the closure defect interpreted as a local measure of the curvature inside the coarse-grained region. It is nevertheless possible to cancel the closure defect by a Lorentz boost. We further identify a Lorentz-invariant observable related to the area and closure defect, which we name ``rest area''. Its physical meaning remains an open issue.

\end{abstract}

\maketitle

\tableofcontents



\vspace*{10mm}
In the context of loop quantum gravity \cite{lqg-review1,lqg-review2,lqg-review3}, the quantum states of geometry are defined as spin network wave-functions. Those are the quantization of classical spinor networks \cite{twisted1,spinor,spinor_johannes, spinor_holo} and are graphs dressed with spins on the edges and intertwiners on the vertices. Spins are integers labeling the irreducible representations of the $\SU(2)$ Lie group while intertwiners are $\SU(2)$-invariant states (single states) in the tensor product of the representations living on the edges attached at the considered vertex. These algebraic data define the (quantum) geometry of the three-dimensional spatial slice through the study of geometry observables, such as areas, volumes and angles, raised to quantum operators acting on the Hilbert space of spin network states. The main challenge of the loop quantum gravity program is then to define the dynamics of those state and understand how it reproduces the general relativity dynamics in a (yet-to-be-precisely-defined) large scale and low energy regime.
To this purpose, it is necessary to investigate the deformations of spinor networks and spin network states, in order to identify which ones should be interpreted as generating the space-time diffeomorphisms (thus implementing the quantum gravity's dynamics), and to define a consistent coarse-graining of those states to go from the Planck scale to higher distances.

In the present paper, we focus on  a bounded and connected region of a spinor network or spin network at the quantum level. The objective is double. On the one hand, we would like to understand how a region can be coarse-grained to a single point and what information characterizes its internal structure. On the other hand, we would like to describe the boundary deformations of this region and what information characterizes the geometry of the region's boundary as observable by an outside observer that can not directly probe its interior. These two questions are clearly complementary.

To compare a non-trivial region to a single point, we start by reviewing the data and observables associated to a single vertex of a spin(or) network. Geometrically, a single vertex is interpreted as defining a (convex) polyhedron (in 3d Euclidean space). Using the spinor formalism, a basic set of $\SU(2)$-invariant observables, forming a closed algebra under the Poisson bracket, was identified \cite{spinor, UN}. These observables generate the deformations of the vertex. One further shows that the deformations preserving the total boundary area dual to the vertex (i.e the boundary area of the associated polyhedron) form the unitary group $\U(N)$ where $N$ is equivalently the number of edges attached to the vertex or the number of faces of the polyhedron. All this holds at the quantum level: we can identify the deformation operators acting on the space of $N$-valent intertwiners and define the area-preserving action of $\U(N)$ on these quantum states.

This action of the unitary group is at the heart of the $\U(N)$ approach to polyhedra and intertwiners \cite{un0,un1,un2,un3,UN}. This is a set of mathematical tools allowing to probe the space of polyhedra and intertwiners, providing exact formulas for the dimension of intertwiner spaces \cite{un1,bhentropy_danny} and thus for the black hole entropy in loop quantum gravity (see \cite{bh_karim1,bh_karim2,bh_kaul_review,bh_jacobo_review} for the description of quantum isolated horizons in terms of $\SU(2)$ intertwiners), allowing to define coherent intertwiner states interpreted as semi-classical polyhedra \cite{un2}, leading to a proposal of holomorphic simplicity constraints for spinfoam models for quantum gravity \cite{un3,un4,un4_conf}, allowing to project onto  a cosmological sector of loop quantum gravity by imposing invariance under $\U(N)$ \cite{spinor,2vertex,2vertex_conf}, and finally providing operators allowing to translate between recursion relations for spinfoam amplitudes and the Hamiltonian constraints of the canonical theory \cite{recursion,recursion3d,recursion4d,recursion_spinor,SFcosmo_merce}.
These applications hint towards the $\U(N)$ structure being an essential feature and tool in the study of loop quantum gravity.

Here, we aim at generalizing this analysis of a single vertex in loop gravity to an an arbitrary closed region of a spin network: identifying the $\SU(2)$-invariant observables and the deformations of the region and investigate if the action of $\U(N)$ transformations can be extended to that case. Understanding and classifying the deformations of spin(or) networks is a necessary step towards  a discrete and quantum equivalent to space-time diffeomorphisms in loop quantum gravity.

\medskip

This short paper is structured as follows. The first section reviews the formulation of the loop gravity phase space on a given graph in terms of spinor variables and spinor networks, as introduced in \cite{twisted1,spinor,spinor_johannes,un2,un4,twisted2,spinor_conf}, and its quantization yielding spin network wave-functions for the quantum geometry. The main purpose of this short section is to fix the notations and make the present paper self-contained. The second section analyzes the case of a single vertex in loop gravity, defined as a bunch of spinors satisfying the closure constraints at the classical level, or as a $\SU(2)$-intertwiner at the quantum level. We define the $\SU(2)$-invariant observables and describe the deformations that they generate. We distinguish the transformations that preserve the boundary area from those that don't. We generalize all the usual formulas to the case when the closure constraints are relaxed.

The heart of the paper starts with the third section, that tackles the problem of defining $\SU(2)$-invariant observables for a bounded region of a spin(or) network. We define the region as a connected set of vertices together with all the edges linking them to each other. The issue consists in the $\SU(2)$ gauge transformations at every (internal) vertex. Using a gauge fixing procedure introduced in \cite{noncompact}, we can pull back all the spinor variables and information to a single (arbitrarily-chosen) vertex within the region. This effectively maps the bounded region onto a single vertex with boundary edges plus self-loops . The boundary edges still describe the boundary of the region with the exterior, while the self-loops encode the non-trivial holonomies and curvature within the region. This allows to interpret any bounded region as a single vertex.

The big difference is that the closure constraints is clearly violated. The closure defect is related to the holonomies around the self-loops and can be interpreted as a coarse-grained measure of the region's curvature and internal structure. We describe this in details in section IV, as well as the ambiguities in the gauge-fixing  procedure. Technically, these amount to the choice of a maximal tree in the internal graph. As we will discuss, all the observables that we introduce unfortunately depend on the chosen tree.

In section V, we introduce Lorentz transformations on the spinors that allow to go in and out of the closure constraints. They allow to uniquely map any coarse-grained set of boundary spinors onto a new set satisfying the closure constraints. We identify $\SL(2,\C)$-invariant observables and define a Lorentz-invariant rest area in terms of the closure defect and boundary area. In short, a unique boost sends the closure defect to 0 and the boundary area onto the rest area. The geometrical meaning of this boost as well as the physical relevance of this rest area are open issues. But we believe this should be relevant to the study of the quantum black holes in loop gravity.

Finally, section VI discusses the coarse-graining of the spin network wave-functions and raises the problem of having to deal with mixed state and density matrices. We conclude with the possible applications of our framework.

\section{The Spinor Network Phase Space}
\label{section1}

Working on a fixed (oriented an connected) graph $\Gamma$, the loop gravity phase space is easily described in terms of spinor variables \cite{twisted1,spinor,spinor_johannes}. One introduces spinor variables $z^{v}_{e}\in\C^{2}$ for each edge around each vertex, thus having two spinors $|z^{s,t}_{e}\ra$ for each edge $e$ with $s,t$ short for the source and target vertices $s(e),t(e)$ of the edge. We endow each spinor $z$ with the canonical Poisson bracket:
\be
\{z^{A},\bz^{B}\}=-i\delta^{AB}\,,
\ee
where the indices $A,B=0,1$ label the component of the spinor.

We use a convenient bra-ket notation for the spinors:
\be
|z\ra\,=\,\mat{c}{z^0 \\ z^1} \in\C^2,
\qquad
\la z| \,=\,\mat{cc}{\bz^0 & \bz^1} \in\C^2,
\ee
and define the dual spinor:
\be
|z]
\,=\,
\varsigma\,|z\ra
\,=\,
\eps\,\mat{c}{\bz^0 \\ \bz^1}
\,=\,
\mat{c}{-\bz^1 \\ \bz^0}\,,
\qquad
\eps=\mat{cc}{0& -1 \\ 1 & 0}
\ee
Note that a spinor and its dual define an orthonormal basis of $\C^{2}$ since $\la z|z\ra=[z|z]$ and $[z|z\ra=0$.

We have two $\SU2)$-invariant scalar products on $\SU(2)$:
\be
\la z|w\ra
\,=\,
(\bz^{0}w^{0}+\bz^{1}w^{1}),
\qquad
[ z|w\ra
\,=\,
(z^{0}w^{1}-z^{1}w^{0}).
\ee
The bilinear form $[ z|w\ra$ is moreover holomorphic and further $\SL(2,\C)$-invariant.

The set of spinors on the graph $\Gamma$ satisfy two sets of constraints, the closure constraints and the matching constraints, respectively attached to the graph's vertices and edges:
\begin{itemize}
\item Closure constraints around every vertex $v$:
\be
\cC^a_{v}
\,\equiv\,
\f12\,\sum_{e\ni v} \la z^{v}_{e}|\sigma^a|z^{v}_{e}\ra =0\,,
\ee
where  the index $a$ runs from 1 to 3 and the $\sigma^a$ are the Pauli matrices, normalized such that each of them square to the identity, $(\sigma^a)^2=\id$. This is equivalent to the requirement that the $2\times2$ matrix $\sum_{e\ni v} |z^{v}_{e}\ra\la z^{v}_{e}|$ be proportional to the identity, or more precisely:
\be
\sum_{e\ni v} |z^{v}_{e}\ra\la z^{v}_{e}| =\cA_{v} \,\id\,,
\qquad
\textrm{with}
\quad
\cA_{v}\equiv \f12\sum_{e\ni v} \la z^{v}_{e}|z^{v}_{e}\ra\,.
\ee

\item Matching constraints along each edge $e$:
\be
\cM_{e}
\,\equiv\,
\la z^{t}_{e}|z^{t}_{e}\ra-\la z^{s}_{e}|z^{s}_{e}\ra=0\,.
\ee

\end{itemize}
Their Poisson brackets close and they all together form a first class constraint system:
\be
\{\cM_{e},\cM_{\te}\}=0,
\quad
\{\cM_{e},\cC_{v}\}=0,
\quad
\{\cC_{v}^{a},\cC_{\tv}^{b}\}\,=\,\delta_{v\tv}\,i\eps^{abc}\cC^c_{v}\,.
\ee
They generate well-defined finite gauge transformations on the spinors:
\begin{itemize}
\item $\SU(2)$-rotations around each vertex $v$:
\be
\forall e\ni v,\quad
|z^{v}_{e}\ra
\,\longrightarrow\,
h^{v}\,|z^{v}_{e}\ra,
\qquad
\textrm{with}
\quad
h^{v}\in\SU(2)\,.
\ee

\item $\U(1)$-phase multiplications on each edge $e$:
\be
|z^{s}_{e}\ra
\,\longrightarrow\,
e^{+i\theta_{e}}\,|z^{s}_{e}\ra,
\quad
|z^{t}_{e}\ra
\,\longrightarrow\,
e^{-i\theta_{e}}\,|z^{t}_{e}\ra,
\qquad
\textrm{with}
\quad
e^{\pm i\theta_{e}}\in\U(1)\,.
\ee
\end{itemize}
Thus the spinor phase space on the graph $\Gamma$ is defined as the symplectic quotient $\C^{4E}//(\SU(2)^{V}\times\U(1)^{E})$ of the spinor space $\C^{4E}$ by the closure and matching constraints, that is sets of spinors satisfying these constraints and up to gauge transformations.

\medskip

Gauging out the phase transformations, we recover the standard formulation of the loop gravity phase space in terms of holonomy-flux variables \cite{spinor,spinor_johannes}. To this purpose, we introduce the vectors, as the projection of the spinors on the Pauli matrices:
\be
\vV_{e}^{v}
\,\equiv\,
\f12\la z^{v}_{e}|\vsigma|z^{v}_{e}\ra
=
-\f12 [z^{v}_{e}|\vsigma|z^{v}_{e}]
\quad\in\R^{3}\,,
\qquad
V_{e}^{v}=|\vV_{e}^{v}|=\f12\la z^{v}_{e}|z^{v}_{e}\ra\,.
\ee
and the holonomies, as the $\SU(2)$ group elements that map one spinor at the source of an edge onto the target spinor:
\be
g_{e}
\,\equiv\,
\f{|z^{t}_{e}]\la z^{s}_{e}|-|z^{t}_{e}\ra [z^{s}_{e}|}
{\sqrt{\la z^{t}_{e}|z^{t}_{e}\ra\la z^{s}_{e}|z^{s}_{e}\ra}}
\quad\in\SU(2)\,,
\qquad
g_{e}\,|z^{s}_{e}\ra=|z^{t}_{e}],
\quad
g_{e}\,|z^{s}_{e}]=-|z^{t}_{e}\ra.
\ee
The holonomies are well-defined due to the matching constraint enforcing that the spinors $z^{s}_{e}$ and $z^{t}_{e}$ have the same norm thus ensuring the existence of a unique $\SU(2)$ group element mapping one to the other. Acting as $\SO(3)$ rotations in the adjoint action, they also map the source vectors onto the target vectors (up to a change of orientation):
\be
g_{e}\triangleright \vV_{e}^{s}=- \vV_{e}^{t},
\qquad\textrm{with}\quad
(g\triangleright \vV)\cdot\vsigma=g\triangleright (\vV\cdot\vsigma)= g(\vV\cdot\vsigma)g^{-1}=\vV\cdot g\vsigma g^{-1}
\,.
\ee
Vectors and holonomies are obviously invariant under the $\U(1)$-phase multiplications and they satisfy the expected $T^{*}\SU(2)$ Poisson brackets. Indeed the group elements are (weakly) commutative and the vectors act as the $\su(2)$ Lie algebra elements:
\be
\{g_{e},g_{\te}\}\approx 0,
\quad
\{g_{e},\vV_{e}^{s}\}= \,g_{e}\,\left(\f{+i\vsigma}{2}\right)\,,
\quad
\{g_{e},\vV_{e}^{t}\}=\,\left(\f{-i\vsigma}{2}\right)\,g_{e}\,,
\ee
with all the vectors commuting with each other and individually satisfying the $\su(2)$ commutators:
\be
\{V^{a},V^{b}\}=-\f{i}{2}\eps^{abc}V^{c}\,.
\ee
These holonomy-flux variables parametrize the symplectic quotient by the matching constraints, $\C^{4E}//\U(1)^{E}\,\sim \,(T^{*}\SU(2))^{\times E}$. This is indeed the loop gravity phase space on the graph $\Gamma$ before imposing the invariance under $\SU(2)$ gauge transformations.

So we still have to impose the closure constraints, which now read simpler in terms of the vectors:
\be
\vcC_{v}=\sum_{e\ni v} \vV^{v}_{e}=0\,.
\ee
They generate $\SU(2)$ gauge transformations parametrized by group elements $\{h^{v}\}\in\SU(2)^{\times V}$ acting  on the vectors and holonomies  as:
\be
|z^{v}_{e}\ra\la z^{v}_{e}|
\,\longrightarrow\, 
h^{v}|z^{v}_{e}\ra\la z^{v}_{e}|(h^{v})^{-1}\,,
\qquad
\vV_{e}^{v}
\,\longrightarrow \,
h^{v}\,\triangleright \,\vV_{e}^{v},
\qquad
g_{e}
\,\longrightarrow\,
 h^{t(e)}\,g_{e}\,(h^{s(e)})^{-1}\,.
\ee
In particular, the closure constraints are of course invariant under such gauge transformations.
Gauging out these closure constraints gives us the (gauge-invariant) loop gravity phase space $(T^{*}\SU(2))^{\times E}//\SU(2)^{\times V}$ on the graph $\Gamma$, which is the phase space over the configuration space $\SU(2)^{\times E}/\SU(2)^{\times V}$ defined by the holonomies up to $\SU(2)$ transformations.

Geometrically, the closer constraints $\sum_{e\ni v} \vV^{v}_{e}=0$ around a vertex $v$ means, by Minkowski theorem, that there exists a unique convex polyhedron dual to that vertex, such that each vector $\vV^{v}_{e}$ living on an edge attached to $v$ becomes the normal vector to a face of that polyhedron, i.e. its direction is orthogonal to the face's plane and its norm gives the area of the face. The interested reader can find more details on this geometrical interpretation and the polyhedron reconstruction in \cite{polyhedron}.
Gluing those polyhedra together by the matching constraints, which equate the area of the faces without necessarily ensuring the matching of the face shape, defines {\it twisted geometries} as the classical kinematical arena of loop gravity \cite{twisted1}. These are a generalization of Regge geometries allowing for torsion and extrinsic curvature (see \cite{twisted_cpt} for a discussion). 
These discrete twisted geometries can further be thought of as a discretization -or more correctly a sampling- of continuous geometries defined in terms of  connection and triad fields \cite{marc,spinning}.

\medskip


The standard loop-quantization is to choose wave-functions of the group elements $g_{e}$ along the edges of the graph, $\phi_{\Gamma}(\{g_{e}\})$. Matrix elements of the $g_{e}$'s act by multiplication, while vectors act as the Hermitian generators $\vJ$ of the $\su(2)$-algebra defined as differential operators on the wave-functions:
\be
\widehat{\vV^{s}_{e}}\,\phi_{\Gamma}(g_{e},g_{\te},..)
\,=\,
-\phi_{\Gamma}(g_{e}\vJ,g_{\te},..),\qquad
\widehat{\vV^{t}_{e}}\,\phi_{\Gamma}(g_{e},g_{\te},..)
\,=\,
\phi_{\Gamma}(\vJ g_{e},g_{\te},..)\,.
\ee
This reproduces the commutators of the $T^{*}\SU(2)$ algebra. Then imposing the closure constraints amounts to requiring the $\SU(2)$ gauge-invariance  of the wave-functions:
\be
\phi_{\Gamma}(\{g_{e}\})
\,=\,
\phi_{\Gamma}(\{ h^{t(e)}\,g_{e}\,(h^{s(e)})^{-1}\}),
\qquad
\forall h^{v}\in\SU(2)^{\times V}.
\ee
Choosing as scalar product the integration over the group elements $g_{e}$ with  the  $\SU(2)$ Haar measure , $\la \phi |\tilde{\phi} \ra\equiv\int_{\SU(2)^{E}}\bar{\phi} \tilde{\phi}$ , this defines the Hilbert space of loop gravity wave-functions on the graph $\Gamma$ as the space of gauge-invariant squared-integrable functions $L^{2}(\SU(2)^{\times E}/\SU(2)^{\times V})$. A standard basis of this space is provided by the spin network functions. These are defined by a choice of irreducible $\SU(2)$ representations on each edge $e$, that is a spin $j_{e}\in\N/2$, and an intertwiner $I_{v}$ for each vertex. The intertwiner $I_{v}$ is a $\SU(2)$-invariant tensor mapping the tensor product of incoming irreps to the tensor product of outgoing irreps:
\be
I_{v}:\quad
\bigotimes_{e|v=t(e)}\cV^{j_{e}} 
\,\mapsto\,
\bigotimes_{e|v=s(e)}\cV^{j_{e}}\,,
\ee
where $\cV^{j}$ is the $(2j+1)$-dimensional Hilbert space of the $\SU(2)$ representation of spin $j$. The corresponding spin network function is then defined as the contraction of the matrix elements of the group elements in the appropriate representations with the intertwiners:
\be
\phi_{\Gamma}^{j_{e},I_{v}}(\{g_{e}\})
\,\equiv\,
\tr\, \bigotimes_{e} D^{j_{e}}(g_{e})\times  \bigotimes_{v}I_{v}
\,=\,
\sum_{m^{s,t}_{e}}
\prod_{e}\la j_{e}m^{t}_{e} | g_{e}| j_{e}m^{s}_{e}\ra \,
\prod_{v}\bigotimes_{e|v=s(e)} \la j_{e}m^{s}_{e} | I_{v} \bigotimes_{e|v=t(e)} |j_{e}m^{t}_{e}\ra\,.
\ee
The scalar product between such functions is given by the Peter-Weyl theorem ensuring the orthogonality of the matrix elements with respect to the Haar measure on $\SU(2)$:
\be
\la \phi_{\Gamma}^{j_{e},I_{v}}|\phi_{\Gamma}^{\tilde{j}_{e},\tilde{I}_{v}}\ra
\,=\,
\prod_{e}\f{\delta_{j_{e},\tilde{j}_{e}}}{2j_{e}+1}\,\prod_{v} \la I_{v}| \tilde{I}_{v}\ra\,.
\ee

\smallskip

Another quantization scheme is to start with the spinor phase space and canonically quantize it as proposed in \cite{spinor}.
The quantization of this phase space is straightforward. One raises the two components of all the spinor variables $z^A$ to annihilation operators for harmonic oscillators, while their complex conjugate become the corresponding creation operators:
\be
z^A \,\longrightarrow\, a^A,\quad
\bz^A \,\longrightarrow\, a^A{}^\dagger,\qquad
[a^A,a^B{}^\dagger]=\delta^{AB}\,.
\ee
The vectors and holonomies now become composite operators in the $a$'s and $a^{\dagger}$'s \cite{spinor,spinor_johannes,spinor_holo}. Considering a single spinor $z$ and dropping the indices $v$ and $e$ for now, this gives us  Schwinger representation for the $\su(2)$ Lie algebra:
\be
V^z\,\longrightarrow\,J^z\,\equiv\,\f12 \big{(}a^0{}^\dagger a^0- a^1{}^\dagger a^1\big{)},\quad
V^\pm\,\longrightarrow\,
\left|\begin{array}{l}
J^+\,\equiv\, a^0{}^\dagger a^1\\
J^-\,\equiv\, a^1{}^\dagger a^0
\end{array}\right.,\qquad
V\,\longrightarrow\, E\,\equiv\,\f12 \sum_A  a^A{}^\dagger a^A,
\ee
\be
[J^z,J^\pm]=\pm J^\pm_i,\quad
[J^+,J^-]=2J^z\,\qquad
[E,\vJ]=0\,.
\ee
The $\vJ$'s are the $\su(2)$ generators while $E$ is the Casimir operator giving the spin and satisfying $\vJ^{2}=E(E+1)$. This allows to generate all the irreducible representations of $\SU(2)$ with arbitrary spin $j$ from the tensor product of two copies of the harmonic oscillator Hilbert space:
\be
\cH_{HO}\otimes\cH_{HO}
\,=\,
\bigoplus_{j\in\N/2} \cV^j,
\qquad
|j,m\ra=|n_0,n_1\ra_{HO}
\quad\textrm{with}\,\,
n_{0,1}=j\pm m\,.
\ee
This scheme naturally leads to a Bargmann representation of the loop gravity Hilbert space in terms of holomorphic functions in the spinors $\phi(\{z_{e}^{v}\})$ provided with the Gaussian measure \cite{spinor,spinor_johannes,spinor_holo}. We further impose that these wave functions be invariant under both $\U(1)$ and $\SU(2)$ gauge transformations.  Then both holonomy and vector operators are constructed as differential operators in the $z$'s \cite{spinor_holo}. Although this is clearly a different choice of polarization from the standard quantization scheme (indeed the $g$'s are not holomorphic in the $z$'s), it is shown to be unitarily-equivalent \cite{spinor_johannes}.

This equivalence is best illustrated by the definition of coherent spin network states as wave-functions of the holonomies but labeled by the spinors \cite{un3,un4,generating}:
\be
\phi_{\Gamma}^{z_{e}^{v}}(\{g_{e}\})
\,\equiv\,
\int [dh^{v}] \, e^{\sum_{e}[z_{e}^{t}|h^{t(e)}g_{e}(h^{s(e)})^{-1}|z_{e}^{s}\ra}\,.
\ee
These are holomorphic in the $z$'s and can be decomposed in the spins $j_{e}$ by simply expanding the exponentials. They provide an over-complete basis of the loop gravity Hilbert space and are peaked on the classical spinor networks defined by the spinor labels. They are thus interpreted as defining semi-classical spinor networks or twisted geometries.

\section{Deformations of a Single Vertex}
\label{1vertex}

Our goal is to describe the (gauge-invariant) degrees of freedom associated to a bounded region of a spinor network at the classical level and spin network at the quantum level.
We call a bounded region as a connected set of vertices, together with all the edges linking them to each other, plus all the edges linking these considered vertices to the outside vertices. We will refer to the region's vertices and the edges between them as the {\it bulk}, while the edges connected to outside vertices are referred to as the {\it boundary}.
Let us point out that such a bounded region is a priori defined combinatorially and does not yet correspond to a bounded region of the actual classical spatial geometry supposed to arise at large scale after a suitable coarse-graining procedure. For instance, we can not ensure that the (spherical) topology of the spatial region's boundary: we might have forgotten in our initial choice of vertices any vertex actually close to all the bulk vertices and a posteriori living inside the considered spatial region. This has to be determined by a careful analysis to the quantum state of geometry of the considered bounded region and its boundary.

Our logic, in the present work, is to appropriately describe the gauge-invariant degrees of freedom associated to the bounded region's geometry, to distinguish the inside/bulk degrees of freedom (associated to the geometry fluctuations inside the region) and the boundary degrees of freedom through which the region will interact with the outside. We thus need to compare the considered bounded region it to its coarse-grained version as a single vertex, analyze the mathematical differences between the two objects and determine in which capacity does the region have an internal structure and which are the extra degrees of freedom associated to it.

To this purpose, we first review in this section the single vertex case, with no bulk structure. We will study in detail in the next section the phase space and quantum degrees of freedom associated to a region with non-trivial bulk. Below, we define the gauge-invariant observables associated to a single vertex, identify area-preserving deformations as $\U(N)$ transformations and introduce a complete basis of area-changing operators. The interested reader will find extensive details on this in \cite{UN}.

\begin{figure}[h]
\includegraphics[height=5cm]{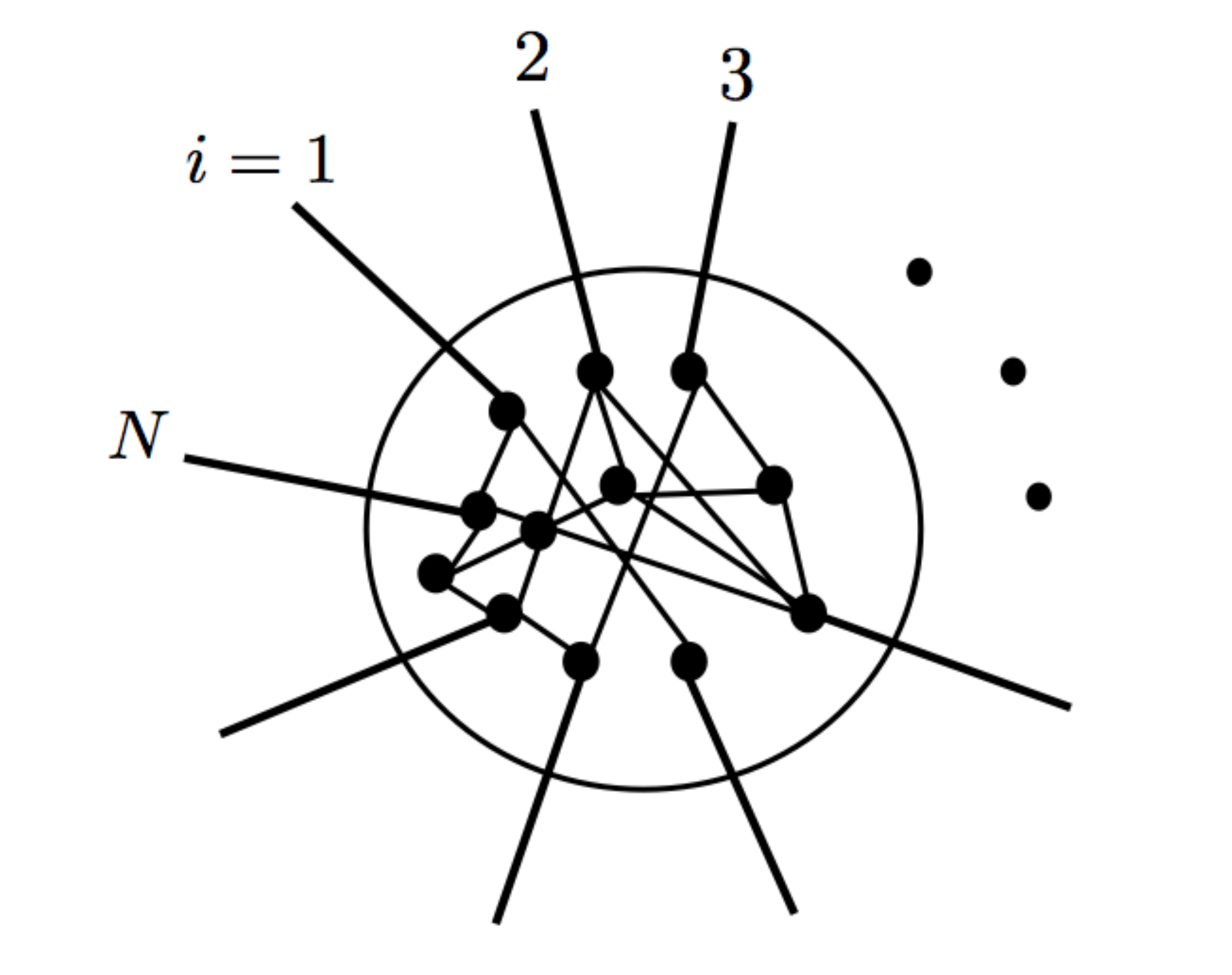}
\caption{\label{fig region} A bounded region of a spin network, with its internal graph and its boundary edges labeled $i=1..N$.}
\end{figure}

\medskip

Considering a single vertex $v$, we drop the index $v$ and label all the edges $e$ attached to that vertex by $i$ running from 1 to $N$. The spinor phase space restricted to the variables attached to that vertex is $\C^{2N}$ provided with the canonical Poisson bracket:
\be
\{z^A_i,\bz^B_j\}=\,-i\delta^{AB}\delta_{ij}\,,
\qquad
\{z^A_i,z^B_j\}=\{\bz^A_i,\bz^B_j\}=0\,,
\ee
together with the closure constraints on these spinors:
\be
\vcC
\,=\,
\f12\sum_i \la z_i|\vsigma|z_i\ra =0\,.
\ee
These constraints are first class and generate global $\SU(2)$ gauge transformations on the $N$ spinors. Vectors and areas are defined as before:
\be
\vV_{i}=\f12\,\la z_i|\vsigma|z_i\ra,
\qquad
V_{i}=\f12\, \la z_i|z_i\ra,
\qquad
\cA=\sum_{i} V_{i}=\f12\sum_i \la z_i|z_i\ra\,,
\ee
with $\cA$ giving the total boundary area of the polyhedron dual to the vertex.

A basis of gauge invariant observables, thus commuting with the closure constraints, is provided by the $E$ and $F$ observes defined as the scalar products between the spinors\cite{un0,un1,un2,un3,un4,UN}:
\be
E_{ij}=\la z_i|z_j\ra=\bE_{ji},
\qquad
F_{ij}=[ z_i|z_j\ra=-[ z_j|z_i\ra=-F_{ji},
\qquad
\bF_{ij}=\la z_j|z_i]=-\la z_i|z_j]\,.
\ee
Both sets of quadratic functions in the spinors are $\SU(2)$-invariant. However, the $F_{ij}$'s are holomorphic in the $z$'s and are further invariant under global $\SL(2,\C)$ transformations: they label the orbits of spinors in $\C^{2N}$ under $\SL(2,\C)$  \cite{un4,UN}.

We can express the original vector scalar products from these spinor scalar products:
\be
\vV_i\cdot\vV_j
\,=\,
2E_{ij}E_{ji}-E_{ii}E_{jj}
\,=\,
-2\bF_{ij}F_{ij}+E_{ii}E_{jj}\,.
\ee
But the important points is that these $\SU(2)$-invariant observables form a closed algebra \cite{un1,un2}:
\beq
\label{algebra}
{\{}E_{ij},E_{kl}\}&=&
-i\left(\delta_{jk}E_{il}-\delta_{il}E_{kj} \right)\\
{\{}E_{ij},F_{kl}\} &=& -i\left(\delta_{il}F_{jk}-\delta_{ik}F_{jl}\right),\qquad
{\{}E_{ij},\bF_{kl}\} = -i\left(\delta_{jk}\bF_{il}-\delta_{jl}\bF_{ik}\right),\nn \\
{\{} F_{ij},\bF_{kl}\}&=& -i\left(\delta_{ik}E_{lj}-\delta_{il}E_{kj} -\delta_{jk}E_{li}+\delta_{jl}E_{ki}\right), \nn\\
{\{} F_{ij},F_{kl}\} &=& 0,\qquad {\{} \bF_{ij},\bF_{kl}\} =0.\nn
\eeq
%
By introducing $E_\rho\equiv\sum_{i,j}\rho_{ij}E_{ij}$ and $F_m\equiv\sum m_{ij}F_{ij}$ for a Hermitian matrix $\rho$ and an antisymmetric matrix $m$, we can write those brackets in a more compact fashion:
$$
{\{}E_\rho,F_m\} = +iF_{m\rho+{}^t\rho m}\,, \qquad
{\{}E_\rho,\bF_m\} =-i\bF_{\bar{\rho}m+m\rho^\dagger}\,,\qquad
{\{} F_m,\bF_m\}= -4iE_{m^\dagger m}.
$$
%

In particular, the $E_{ij}$ form a $\u(N)$-algebra \cite{un0}. This $\u(N)$ structure is fundamental from the mathematical perspective, it is crucial in understanding the structure of $\SU(2)$ intertwiners at the quantum level (see \cite{un1,un2,un3} for details and \cite{UN} for a full review). The $E$-observables Poisson-commute with the total boundary area, 
$\{E_{ij},\cA\}=0$, and generate $\U(N)$-transformations on the spinors that preserve the area $\cA$. Indeed, considering a Hermitian matrix $\rho$,
one computes \cite{un2,un4}:
\be
\{E_{ij},|z_k\ra\}
=\,i\delta_{ik}\,|z_j\ra,
\quad
\{E_\rho,|z_k\ra\}
=\,i\sum_j\rho_{kj}\,|z_j\ra,
\quad
e^{\{E_\rho,\cdot\}}\,|z_k\ra
= \sum_j(e^{i\rho})_{kj} \,|z_j\ra,\qquad
\textrm{where}\quad
e^{i\rho}\in\U(N)\,.
\ee
One checks a posteriori that these $\U(N)$ transformations preserve both the closure constraints and the total area,
\be
\{z_i\}_{i=1..N}
\,\overset{\U(N)}\longrightarrow\,
\{\tz_i\}_{i=1..N}
=U\vartriangleright\{z_i\}_{i=1..N}
=\{\sum_j U_{ij}z_j\}_{i=1..N}
\qquad\textrm{for}\quad
U\in\U(N)\,,
\ee
by computing the transformation of  the $2\times 2$ matrix $\cX\,\equiv\, \sum_i |z_i\ra\la z_i|$:
\be
\sum_i |\tz_i\ra\la \tz_i|
=\sum_{ijk} U_{ij}\overline{U_{ik}}\,|z_j\ra\la z_k|
=\sum_{jk} \delta_{jk}\,|z_j\ra\la z_k|=\sum_i |z_i\ra\la z_i|\,. \nn
\ee
Let us underline that the $\U(N)$ action preserves the whole matrix $\cX$ even when the closure constraints $\vcC=\f12\tr \cX\vsigma=0$ are relaxed.
An important equation is the {\it Casimir equation} expressing the $\u(N)$ Casimir in terms of the boundary area and the closure vector:
\be
\label{casimirE}
\cE^{2}\,\equiv\,
\f12\sum_{i,j} E_{ij}E_{ji}
\,=\,
\f12\sum_{i,j} |\la z_{i}|z_{j}\ra|^{2}
\,=\,
\f12\tr \cX^{2}
\,=\,
\cA^{2}+|\vcC|^{2},
\qquad
\cX= \sum_i |z_i\ra\la z_i|
\,=\,\cA\id+\vcC\cdot\vsigma\,.
\ee
The area is to be understood as the Casimir of the $\U(1)$ phase transformations (generated by  $\{\cA,\cdot\}$) while the norm of the closure vector gives the $\SU(2)$ Casimir. When considering only closed configurations of spinors, the $\U(N)$ Casimir is entirely determined by the boundary area $\cA$. In the rest of the paper, we will refer to $|\vcC|$ (or to $|\vcC|^{2}$ when there is no ambiguity) as the {\it closure defect}.
It is this equation that will determine at the quantum level how to decompose the space of $\SU(2)$ intertwiners in terms of the irreducible $\U(N)$-representations.

Now turning to the observables $F$ and $\bF$, they will not only deform the dual polyhedron but also change the total area:
\be
\{F_{ij},\cA\}=-2iF_{ij},\qquad
\{\bF_{ij},\cA\}=+2i\bF_{ij}\,.
\ee
We can easily exponentiate those actions since the $F$'s commute with each other and the $\bF$ do the same. This can be traced to the fact that the are holomorphic in the spinor variables while the $\bF$ are purely anti-holomorphic. It is more relevant to look at the action generated by real combinations of $F$'s, thus mixing both $F$'s and $\bF$'s.
%
%
%
We compute the action of $F_m+\bF_m$ on the spinors\footnotemark:
\be
\{F_m+\bF_m,|z_k\ra\}=-im_{kj}|z_j],\qquad
\{F_m+\bF_m,|z_k]\}=-im_{kj}|z_j\ra\,,
\ee
\footnotetext{
One could also consider the action of the other combination $i(F_m-\bF_m)$ but this observable is simply $F_{im}+\bF_{im}$.
}
where the sum over the index $j$ is implicit. We see that it mixes the spinors $|z\ra$'s with their dual. It thus seems more convenient to introduce left and right spinors:
\be
|z^\pm_k\ra\,\equiv\,|z_k\ra\pm|z_k]\,.
\ee
Then the  action of $(F_m+\bF_m)$ simply reads:
\be
\{F_m+\bF_m,|z_k^\pm\ra\}=\,\mp i  m_{kj}|z_j^\pm\ra\,,
\ee
which is obvious to exponentiate:
\be
e^{\{F_m+\bF_m,\cdot\}}\,|z_k^\pm\ra
\,=\,
|z_k^\pm\ra+\mp i m_{kj}|z_j^\pm\ra + \f{(\mp i)^2}{2}m_{kj}^2|z_j^\pm\ra+\dots
\,=\,
(e^{\mp im})_{kj}\,|z_j^\pm\ra\,.
\ee
Let us underline the fact that $m$ is not Hermitian but complex and anti-symmetric. Decomposing it into real and imaginary parts gives us respectively its anti-Hermitian and Hermitian components. Thus the matrix $e^{\mp im}$ is not unitary, but will be in general the product of a unitary matrix and a Hermitian matrix. In particular, as expected, the exponentiated action of $\{F_m+\bF_m,\cdot\}$ does not conserve the total area $\cA$:
\be
\cA=\f14\sum_i\la z_i^+|z_i^+\ra
\,\longrightarrow\,
\widetilde{\cA}=\f14\sum_{j,k} \la z_j^+|z_k^+\ra\,(e^{-i\overline{m}}e^{- im})_{jk}\,.
\ee
Looking back at the commutators of $F$ and $\bF$ with the total boundary area $\cA$, we see that we can interpret the $F$'s as generating the shrinking of the polyhedron while the $\bF$'s expands it. Finally, the $E$'s and $F$'s and $\bF$'s all together generate all the deformations of the single vertex and its dual polyhedron.

We can define a second $\U(N)$ Casimir from the $F$'s observables, but it turns out to be entirely determined by the Casimir $\cE$ introduced above:
\be
\label{casimirF}
\cF^{2}\,\equiv\,
\f12\sum_{i,j}|F_{ij}|^{2}
\,=\,
\f12\tr \cX(2\cA\id-\cX)
\,=\,
2\cA^{2}-\cE^{2}
\,=\,
\cA^{2}-|\vcC|^{2}\,.
\ee
The interest for $\cF$ compared to $\cE$ is nevertheless that it is $\SL(2,\C)$ invariant and not only invariant under $\SU(2)$. This will be particular relevant when relaxing the closure constraints in section \ref{boost}.

\medskip

All this is straightforwardly translated to the quantum level. Quantizing the spinor components $z_{i}^{A}$ as harmonic oscillators $a_{i}^{A}$ and using the normal ordering, we define the corresponding  operators:
\be
\hE_{ij}=\sum_A a_i^\dagger a_j,\qquad
\hF_{ij}=a_i^0 a_j^1- a_i^1 a_j^0,\qquad
\hF^\dagger_{ij}=a_i^0{}^\dagger a_j^1{}^\dagger- a_i^1{}^\dagger a_j^0{}^\dagger\,.
\ee
These are still $\SU(2)$ invariant. Their commutators are quantized without anomaly and reproduce the same algebra as given by eqn.\eqref{algebra}. All the details can be found in \cite{un2,un3,spinor}. The main point is that the Hilbert space of intertwiner states still carries a $\U(N)$-action generated by the $\hE$'s. Actually the intertwiner space for fixed total area provides an irreducible representation of $\U(N)$ \cite{un1} and we get the following ladder structure:
\be
\cH_N=\bigoplus_{J\in\N} \cR^J_N
\qquad\textrm{with}\quad
\cR^J_N=\bigoplus_{\sum_{i=1}^N j_i =J}\textrm{Inv}_{\SU(2)} \cV^{j_1}\otimes..\otimes \cV^{j_N}\,.
\ee
Each Hilbert space $\cR^J_N$ carries an irreducible representation of $\U(N)$, given by the Young tableaux with two horizontal lines made with a equal number of $J$ boxes. The diagonal $\u(N)$-generators $\hat{E}_{ii}$ give twice the value of  the spin $j_i$ living on the $i$-th leg of the intertwiner. The generator of the global phase transformation $\hat{\cA}=\f12\sum_i\hat{E}_{ii}$ gives the total boundary area $J=\sum_{i}j_{i}$.

On the one hand, the $\hE_{ij}$ operators acts on each space $\cR^J_N$ without changing $J$. They describe the deformations of the intertwiner at fixed boundary area. They  generate $\U(N)$ transformations on $\cR^J_N$. When acting on the highest weight vector given by the bivalent intertwiner ($j_{1}=j_{2}=\f J2$, $j_{k\ge 3}=0$), they allow to define coherent intertwiner states with fixed total area $J$ \cite{un2}.
On the other hand, the $\hF_{ij}$ decrease the area $J$ by one while their adjoint $\hF^\dagger_{ij}$ increase the area by one. From this perspective, the $\hF$'s are to be interpreted as annihilation operators while the $\hF^\dagger$'s are creation operators. For instance, we can use the $\hF^\dagger$'s to define coherent intertwiner states \cite{un2} and these turn out to be eigenvectors of the operators $\hF$ \cite{un3,un4}.

Finally, we derive the {\it Casimir equations} at the quantum level\footnotemark:
\be
\widehat{\cE^{2}}\,\equiv\,
\f12\sum_{i,j} \hE_{ij}\hE_{ji}
\,=\,
\hat{\cA}(\hat{\cA}+N-2)+\hat{\cC}^{a}\cdot\hat{\cC}^{a},
\qquad
\widehat{\cF^{2}}\,\equiv\,
\f12\sum_{i,j} \hF_{ij}^{\dagger}\hF_{ij}
\,=\,
\hat{\cA}(\hat{\cA}+1)-\hat{\cC}^{a}\cdot\hat{\cC}^{a},
\ee
\be
\widehat{\cE^{2}}+\widehat{\cF^{2}}
\,=\,
\hat{\cA}(2\hat{\cA}+N-1)\,,
\ee
where we have quantized directly the observables $\cE^{2}$ and $\cF^{2}$ (we haven't defined and quantized their square-root $\cE$ and $\cF$, which would require choosing suitable ordering for non-polynomial observables).
\footnotetext{
We can derive these relations from the formulas for the vector scalar products at the quantum level:
$$
\vJ_{i}\cdot \vJ_{i}
=\f{\hE_{ii}}{2}\left(\f{\hE_{ii}}{2}+1\right),\qquad
\forall i\ne j\,,\quad
\vJ_{i}\cdot \vJ_{j}
=\f12 \hE_{ij}\hE_{ji} -\f{\hE_{ii}}{2}\f{\hE_{jj}}{2}-\f{\hE_{ii}}{2}
=-\f12\hF_{ij}^{\dagger}\hF_{ij} +\f{\hE_{ii}}{2}\f{\hE_{jj}}{2}\,.
$$
}
The closure vector operators $\hat{\cC}^{a}$ are the global $\su(2)$ generators and their Casimir takes the standard discrete values $\cC^2\equiv\hat{\cC}^{a}\cdot\hat{\cC}^{a}=c(c+1)$ with a global half-integer spin $c\in\N$. The case $c=0$ corresponds to intertwiner states. When $c$ does not vanish, we have a quantized closure defect. At fixed total area $J$ and closure defect $c$, we have a $\U(N)$ irreducible representations corresponding to intertwiners states between the $N$ boundary edges and a fictious extra link carrying the closure defect:
\be
\cR^{J,c}_N=\bigoplus_{\sum_{i=1}^N j_i =J}\textrm{Inv}_{\SU(2)} \cV^{j_1}\otimes..\otimes \cV^{j_N}\otimes \cV^c\,.
\ee
This defines the $\U(N)$ irrep with Young tableaux given by two horizon lines made of a different number of boxes, $(J+c)$ for the first one and $(J-c)$ for the second.

\smallskip

This concludes our review and analysis of the $\SU(2)$ observables and deformation operators for a single vertex and intertwiner.  The purpose of the present paper is now to discuss the application of this framework to the definition of deformation operators beyond the single intertwiner for non-trivial regions of arbitrary spin networks.

\section{Coarse-Graining through Gauge Fixing}


Similarly to the case of a single vertex, we would like to study the degrees of freedom associated to a bounded (connected region) of a spinor network (and then of a spin network at the quantum level) having in mind the coarse-graining of loop gravity's kinematics and dynamics. The aim is to understand which degrees of freedom are relevant to the internal structure of the region's geometry and which degrees of freedom will interact with the outside, in order to understand which data are relevant when we coarse-grain the considered region to a single point. This is crucial in the perspective of analyzing the renormalization flow of loop quantum gravity and how the dynamics of the geometry changes with the length scale. It is also important from the point of view of (quantum) black holes, which admit descriptions as point-like particles or single intertwiners in loop quantum gravity although representing highly dense and extended region of space(-time).  It is thus essential to understand in which extent we can effectively describe a bounded region of a spinor network as a single vertex (of a coarse-grained network), with possibly extra-data reflecting the internal structure and curvature of the region's bulk geometry. 

\smallskip

Considering a bounded region of a spinor network, with a possibly complex internal graph and many boundary edges, the spinor variables encode all the information about the (twisted) geometry of the region corresponding to the discretization of the connection and triad fields (for the discrete-continuum correspondence, we refer the interested reader to \cite{marc,spinning}). However this information is redundant due to the $\SU(2)$ gauge-invariance. It would be even further redundant if taking into account the Hamiltonian constraints and the invariance under (quantum) space-time diffeomorphisms, but we focus here on studying the kinematic structures of loop gravity and geometry states not necessarily satisfying the (quantum) Einstein equations. Moreover it is not clear whether the Hamiltonian constraints can be formulated on a fixed graph or if they require to consider more generally superpositions of spin network states living on different graphs. Thus taking into account the $\SU(2)$ gauge-invariance, we need to identify the gauge-invariant information reflecting the bulk geometry deformations of the considered region and its boundary data reflecting how it interacts with (or glues itself with) the outside.

We take the point of view of an external observer who does not have access to the internal graph structure of the region, but can at best access and make measurements on its boundary. It should still be able to feel the internal structure through the non-trivial parallel transport along curves going across that region, and thus access some coarse-grained data about the bulk curvature inside this region. It is possible to formalize this in mathematical terms using a gauge-fixing procedure, introduced in \cite{noncompact}, which trivializes as many holonomies as possible along internal edges while retaining the information about the connection and bulk curvature as  holonomies around closed internal loops. This procedure maps the possibly complicated internal graph onto a flower graph with a single vertex, and as many petals as independent internal loops, thus effectively reducing the bounded region to a single vertex without losing any algebraic information accessible to the external observer but nevertheless discarding the combinatorial information about the initial internal graph.

\smallskip

We choose a bounded region on the oriented graph $\Gamma$ as a connected set of vertices together with all the edges connecting those vertices with each other. We refer to this set of vertices and internal edges as the internal graph $\Gamma_{in}$. We write $V$ and $E$ respectively for the number of internal vertices and edges. The boundary edges are the edges connecting the chosen vertices with an outside vertex. We write $N$ for the number of boundary edges.

The gauge-fixing procedure, introduced in \cite{noncompact} and adapted to the present setting, starts by choosing a maximal tree $T$ in the internal graph, that is a set of edges in $\Gamma_{in}$ going through all the internal vertices and never forming a loop. Such a choice always exists (but never unique except in trivial cases when the internal graph is itself a tree) and consists in $(V-1)$ edges. The important property resulting from the choice of such a maximal tree is the existence of a unique path within $\Gamma_{in}$ along the tree $T$ linking any two internal vertices. We now choose a reference internal vertex $v_{0}$ and write $\cP_{v}^{T}$ or $[v_{0}\arr v]$ for the unique path linking the reference vertex $v_{0}$ to an arbitrary internal vertex $v$. We will omit the subscript $T$ referring to the tree whenever there is no ambiguity. Then we define the holonomy from $v_{0}$ to $v$ along the tree $T$ as the oriented and ordered product of holonomies along the edges of the path  $\cP_{v}^{T}$:
\be
G_{v}
\,\equiv\,
\overrightarrow{\prod_{e\in\cP^{v}}} g_{e}
\,=\,
 g_{v_{n-1}\arr v}^{\eps_{n}}..g_{v_{0}\arr v_{1}}^{\eps_{1}}\,,
\ee
where  $\cP_{v}^{T}$ goes through $(n+1)$ vertices, $v_{0}\arr v_{1}\arr..\arr v_{n-1}\arr v$, and the signs $\eps_{i}=\pm$ register the relative orientation of the considered edge with the path orientation. 

\begin{figure}[h]
\includegraphics[height=5cm]{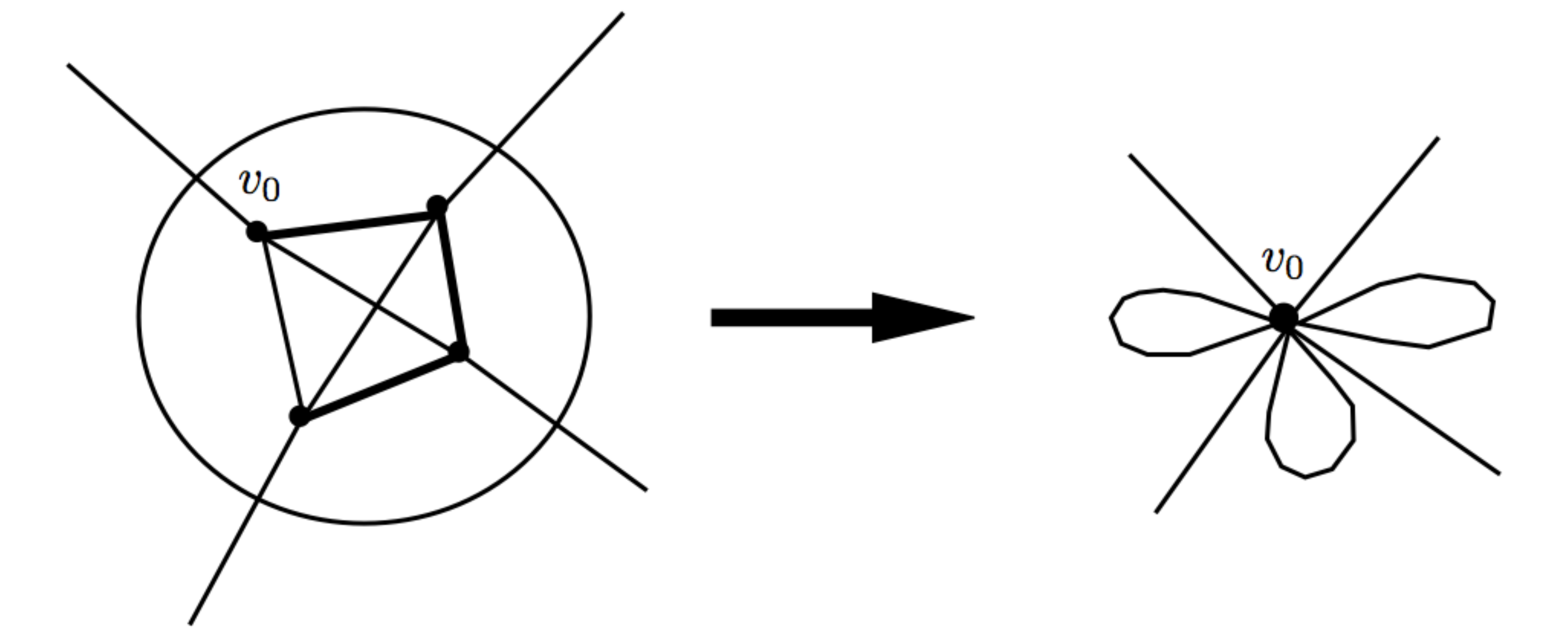}
\caption{\label{fig-cg1} The gauge-fixing of the internal graph to a flower graph: starting with $V=4$ internal vertices linked with $E=6$ internal edges, we choose a reference vertex $v_{0}$ and a tree $T$ in {\bf bold}; we obtain a flower with $E-V+1=3$ petals and the original 4 boundary edges after contracting the internal graph and setting the holonomies to $\id$ along the tree $T$.}
\end{figure}

As we saw earlier, $\SU(2)$ gauge transformations, $h^{v}\in\SU(2)^{\times V}$ act on the spinors as $|z_{e}^{v}\ra\arr  \, |\tz^{v}_{e}\ra=
h^{v}\,|z_{e}^{v}\ra $ and resultingly on the holonomies as $g_{e}\arr\,\tg_{e}=h^{t(e)}g_{e} (h^{s(e)})^{-1}$. We define parallel-transported spinors:
\be
|Z^{v}_{e}\ra
\,\equiv\,
G_{v}^{-1}\,|z^{v}_{e}\ra\,,
\ee
which amounts to a gauge transformation with $h^{v}=G_{v}^{-1}$ somehow pulling back all the spinors back to the reference vertex $v^{0}$. This gauge transformation maps all the group elements along edges of the tree $T$ to the identity:
\be
g_{e\in T}
\,\arr\,
\tg_{e}=\id\,.
\ee
This gauge-fixing trivializes the parallel transport along the tree $T$ and  allows to somehow  synchronize  frames associated to every internal vertex with the chosen reference vertex $v_{0}$.
Every edge  $e\notin T$ not in the tree $T$ defines a loop from the reference vertex $v_{0}$ to the edge $e$ and then back to the $v_{0}$ along the tree $T$. The holonomy associated to that edge $g_{e\notin T}$ is mapped by our our gauge transformation to the holonomy around that loop:
\be
g_{e\notin T}
\,\arr\,\tg_{e\notin T}=G_{t(e)}^{-1}g_{e}G_{s(e)}\equiv \cG_{e}^{T}
\ee 
Then, after contraction of the tree $T$ to $v_{0}$, we are left with a single vertex $v_{0}$ and a remaining single $\SU(2)$-gauge invariance, and a flower graph with $(E-V+1)$ petals corresponding to the independent internal loops associated to every edge not in the tree, as illustrated on fig.\ref{fig-cg1}.

We can consider such a gauge-fixing as the first step of the coarse-graining of the considered bounded region, with the holonomies $\tg_{e\notin T}=G_{t(e)}^{-1}g_{e}G_{s(e)}$ around the independent internal loops as the bulk degrees of freedom. The fully gauge-invariant data is the orbit of these holonomies under the global $\SU(2)$ adjoint action, $\{\tg_{e}\}_{e\notin T}\arr \{h\tg_{e}h^{-1}\}$. This realizes the isomorphism between the configuration space $\SU(2)^{\times E}/\SU(2)^{\times V}$ of the group elements on the internal graph with the reduced flower configuration space $\SU(2)^{\times E-V+1}/Ad\,\SU(2)$.
Then the Poisson brackets with these loop holonomies, or the corresponding loop holonomy operators at the quantum level, will generate the deformations of the internal geometry of the region.

\smallskip

To summarize, if the internal graph of the bounded region is already a tree, i.e. contains no loop, then the whole region can be considered as having non internal structure per se and can be identified  to a single vertex for the external observer without loss of information. Indeed, without internal loops, there can not be any gauge-invariant fluctuations of the connection inside the region and thus no non-trivial curvature can develop itself. Morally, physical degrees of freedom of loop gravity live on loops of the spin networks when there is no boundary and they are for instance the only relevant contributions to the region's entropy (see e.g. \cite{danny-cg, danny-bulk, danny-bf}). Mathematically, when the internal graph is a tree, one can map by a gauge transformation all the holonomies within the region to the identity, thus somehow projecting all the information about the region's geometry onto its boundary. This is an obvious realization of the holographic principle, at the kinematical level.

When there are internal loops and thus petals to the gauge-fixed flower, the region can develop curvature and non-trivial holonomies around the loops. At the kinematical level, we therefore have internal information to retain about the region wen coarse-graining it. At the dynamical level, when taking into account a proper consistent implementation of the diffeomorphism constraints at the quantum level, these bulk degrees of freedom (local to the region) might get projected out onto the boundary once more, thus providing a non-trivial implementation of the holographic principle in loop quantum gravity. This is however outside the scope of the present analysis.

\smallskip

In the next section, we will focus on the boundary degrees of freedom, that is the spinors living on the boundary edges pulled back to the reference vertex, and use this formalism to define the boundary deformations of the region.

\section{Deformations of a Bounded Region}

\subsection{Boundary observables and $\U(N)$ Transformations}

Let us now focus on the boundary edges around the considered region. We label these by $i=1..N$ and write $v_{i}$ for the internal vertex to which the boundary edge $e_{i}$ is attached. Possibly many of these vertices $v_{i}$ are identical. Following the gauge-fixing procedure introduced in the previous section, we call $G_{i}\equiv G_{v_{i}}$ the holonomy from the reference vertex $v_{0}$ to $v_{i}$ along the chosen maximal tree $T$ for the internal graph. Finally, for a given boundary $e_{i}$, we write $z_{i}$ short for the spinor $z_{e_{i}}^{v_{i}}$ irrespectively that the corresponding internal vertex $v_{i}$ is the source or the target of that edge.

Similarly to the internal edges, we define the spinors parallel-transported back from the boundary to the reference vertex $v_{0}$:
\be
|Z_{i}\ra \,\equiv\,
G_{i}^{-1}\,|z_{i}\ra\,
\ee
Now these spinors all transform homogeneously under $\SU(2)$ gauge transformation, solely depending on the gauge transformation $h\equiv h^{v_{0}}$:
\be
|Z_{i}\ra \,\arr\,
|\tZ_{i}\ra
\,=\,
h^{v_{0}}G_{i}^{-1}(h^{v_{i}})^{-1}\,h^{v_{i}})\,|z_{i}\ra\,=\,
h^{v_{0}}G_{i}^{-1}\,|z_{i}\ra
\,=\,
h\,|Z_{i}\ra\,.
\ee
From this, it is clear that the scalar products between parallel-transported spinors is gauge invariant (while the scalar products between the original spinors $z_{i}$ are not). We therefore introduce the following boundary observables:
\be
E^T_{ij}
=\la Z_i|Z_j\ra
=\la z_i|G_iG_j^{-1}|z_j\ra,\qquad
F^T_{ij}=[ z_i|G_iG_j^{-1}|z_j\ra,\qquad
\bF^T_{ij}
=\la z_j|G_jG_i^{-1}|z_i]\,.
\ee
These are $\SU(2)$-invariant, they do not depend on the choice of reference vertex $v_{0}$ but they still depend a priori on the choice of maximal tree $T$.

Since the holonomies weakly commute with each other (their Poisson brackets vanishes if taking into account the matching constraints) and that the holonomies along internal edges obviously commutes with the spinors attached to boundary edges, the spinors $Z_{i}$ satisfy canonical Poisson brackets and the $E^{T}$ and $F^{T}$ defined above satisfy exactly the same algebra \eqref{algebra} as the spinor scalar products for a single vertex. The observables $E^{T}_{ij}$ and $F^{T}_{ij}$ generate the boundary deformations of our bounded region. 

In particular, the $E^{T}_{ij}$ form a closed $\u(N)$ Lie algebra and generates $\U(N)$ transformations as in the single vertex case, that is for a $N\times N$ unitary matrix $U\in\U(N)$:
\be
|Z_{i}\ra
\,\arr\,
\sum_{j} U_{ij }\,|Z_{j}\ra\,,
\qquad
|z_{i}\ra
\,\arr\,
\sum_{j} U_{ij}\,G_iG_j^{-1}\,|z_{j}\ra\,.
\ee
Let us comment that the $2\times 2$ matrix indices are kept implicit in these equations and do not interfere with the $\U(N)$ action, which mixes the label of boundary edges. The observables $E^{T}_{ij}$  still generate the area-preserving transformations, while the action generated by the observables $F^{T}_{ij}$ and $\bF^{T}_{ij}$ will change the total boundary area.

\smallskip

These observables generate transformations that act non-trivially on the spinor variables in the bulk. Indeed, one can see these as open strings connecting one point on the boundary to another through a curve in the bulk. At the quantum level, one follows the standard procedure to quantize the spinors and holonomies (see e.g. \cite{spinor_holo}). Then due to the holonomy insertions $G_{i}$ and $G_{j}$, the operators $\hE^{T}_{ij}$ (and $\hF^{T}_{ij}$) will induce $\pm\f12$ shifts in the spins living on the internal edges along the tree $T$. In particular, it is obvious that the action of these operators on the spin network states depend on the tree $T$. Nevertheless it is not clear that the tree $T$ is relevant to the outside observer. We will investigate this further below.

\subsection{Computing the $\U(N)$-Casimir and Relaxing the Closure Constraints}

For a single vertex, computing the value of the $\U(N)$-Casimir at the classical and quantum levels allowed to determine which representation of the unitary group we were dealing with. We do the same for our new boundary observables. The difference is that the parallel-transported spinors $Z_{i}$ do not a priori satisfy closure constraints and we obtain generically a non-trivial closure defect:
\be
\vcC^{T}=\sum_{i}\la Z_i|\vsigma|Z_i\ra \ne 0\,.
\ee
Only the norm $|\vcC^{T}|$ of this vector is actually $\SU(2)$-invariant, but it is essential to keep in mind that the gauge invariance does not ensure that the closure vector on the boundary vanishes.

This non-vanishing is due to the non-trivial holonomies within the region. Indeed, when gluing the closure constraints individually satisfied by each internal vertices, we won't be able to keep satisfying the closure constraints for the boundary if we use non-trivial $\SU(2)$ group elements. If the internal graph is a tree, there is no loop and non-trivial holonomy in the bulk and the boundary spinors $Z_{i}$ will automatically satisfy the closure constraints. However, as soon as there is a non-trivial holonomy around an internal loop, we will get a non-trivial closure defect. It is important to keep in mind that this violation of the closure constraints is simply due to a non-trivial curvature within the region and thus reflects the existence of an internal structure to the considered region. From this perspective, the closure defect provides a measure of the curvature in the bulk at a coarse-grained level. 
We will later illustrate this explicitly on a one-loop example in section \ref{1loop}.

Going out of the closure constraints when coarse-graining naturally leads to a big issue about coarse-graining spin(or) networks. Since gauge-fixing is simply a first step toward coarse-graining the considered bounded region to a single point, should we further erase all the internal structure and brutally project the state on a vanishing closure vector or should we accept the closure defect as a relevant physical variable and enlarge the space of spin(or) networks to take it into account or does there exist a natural way to rotate a non-vanishing closure vector back to  the constraint surface without losing further information? This issue, and specially the possible extensions of the space of  spin(or) network suggested but the gauge-fixing procedure and the closure defect, will be discussed further in future work. We will nevertheless study in the next section \ref{boost} the possibility of boosting a closure defect back to 0 by a $\SL(2,\C)$ transformations on the spinors. 

\smallskip

One gets the $\U(N)$ Casimir in terms of the closure defect and the boundary area from the Casimir equation \eqref{casimirE} derived earlier in section \ref{1vertex}:
\be
\cE_{T}^{2}\equiv
\f12\sum_{i,j}E^{T}_{ij}E^{T}_{ji}
=
\f12\tr \left(\sum_{i}G_i^{-1}|z_i\ra\la z_i|G_i\right)^{2}
=
\cA^{2}+|\vcC^{T}|^{2}\,.
\ee
While the total area $\cA=\f12\sum_{i}\la z_i|z_i\ra=\f12\sum_{i}\la Z_i|Z_i\ra$ clearly does not depend on the chosen maximal tree, the closure defects and $\U(N)$ Casimir do. Nevertheless, we always have the inequality, $0\le|\vcC^{T}|\le \cA$ (remember that the Minkowski mapping of vectors satisfying the closure constraints onto a convex polyhedron works in Euclidean 3d space, where the standard triangle inequalities always hold).

At the quantum level, intertwiners lives in $\U(N)$ irreducible representations with Young tableaux given as two horizontal lines of equal length \cite{un1}. Now the $\SU(2)$ Casimir does not vanish and we will get Young tableaux with still two horizontal lines (corresponding to the two components of a spinor) but with the second  line possibly shorter than the first, with the difference reflecting the closure defect \cite{un1}.

\subsection{Changing Tree: what's accessible to the outside observer?}

%
%
%
%
%
%
%
%


All our definitions of gauge-fixing, internal holonomies, parallel-transported boundary spinors and the resulting closure defect crucially depend on the choice of a maximal tree $T$ within the internal graph $\Gamma_{in}$. Let us have a deeper look into this. 

\begin{figure}[h]
\includegraphics[height=5cm]{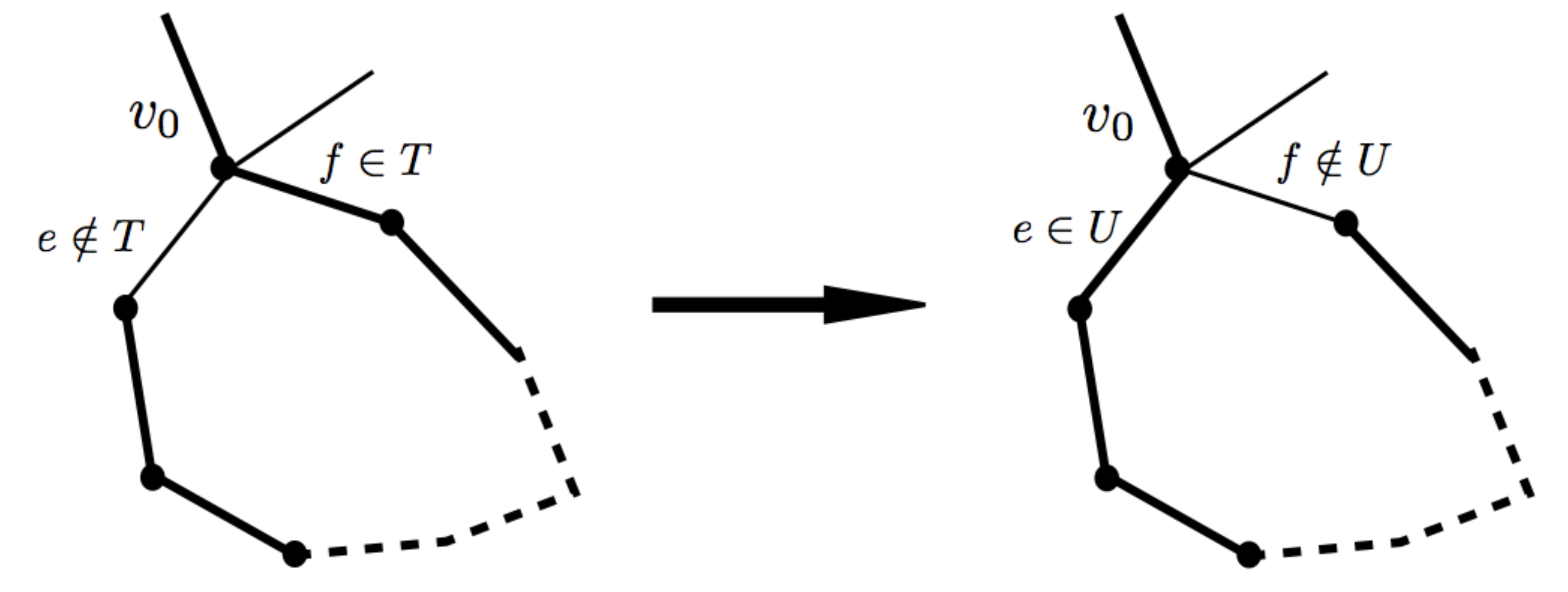}
\caption{\label{fig-tree} An elementary change of tree: starting with the tree $T$ going through the vertex $v_{0}$ with $e\notin T$ and $f\in T$, we exchange the roles of the edges $e$ and $f$ to get a new tree $U$ with  $e\in U$ and $f\notin U$. The edges belonging to the trees are drawn in bold. The loop holonomy $\cG_{e}^{T}$ associated to the edge $e\notin T$ and defined by gauge-fixing along the tree $T$ is equal to the loop holonomy $\cG_{f}^{U}$ associated to the edge $f\notin U$ by gauge-fixing along the tree $U$(up to potentially taking its inverse, depending on the relative orientations of the edges $e$ and $f$ along the loop). It is simply the oriented product of the holonomies along the edges in the loop drawn above.}
\end{figure}

There exist elementary changes of tree, which allows to explore all possible maximal trees on a given graph \cite{noncompact}. Let us choose a vertex to which is attached at least one edge $e$ which is not in the tree $T$. For the sake of simplicity, we will choose that vertex as the reference vertex $v_{0}$, and let us assume that it is the source vertex of $e$. Then there exists a unique path $\cP_{t(e)}^{T}$ along the tree $T$ that links $v_{0}=s(e)$ to the other end of the edge $t(e)$. It goes through a first edge, call it $f\in T$, attached to $v_{0}$. Let us assume for the sake of simplicity that $v_{0}$ is also the source vertex of $f$. We define a new maximal tree $U$ by removing $f$ from the tree and adding $e$ instead, as illustrated on figure \ref{fig-tree}:
\be
e\notin T,\quad f\in T
\qquad\longrightarrow\qquad
e\in U,\quad f\notin U\,.
\ee
In the paths between any two internal vertices along the tree, we need to replace the edge $f$ by the edge $e$ followed by the rest of the loop until $t(f)$. 
This will replace the group element $g_{f}$ by $g_{f}\cG_{e}^{T}$ in all the holonomy formulas for the $G_{v}$ and thus for the $G_{i}$'s between $v_{0}$ and the boundary, where $\cG_{e}^{T}$ is the holonomy around the loop along $T$ associated to the edge $e$. We will also replace the loop to the edge $e\notin T$ by the loop to the edge $f\notin U$ with the holonomy $\cG_{f}^{U}=(\cG_{e}^{T})^{-1}$ (we get the inverse because the edges $e$ and $f$ have opposite orientation around the loop).

Changing the tree changes the paths along which we compute the holonomies to pull back the boundary spinors to the reference vertex. It therefore changes the definition of these parallel-transported boundary spinors. As we see above, it modifies the group elements $G_{i}$'s by some insertion of the internal loop holonomies. In the meanwhile, those holonomies  around internal loops are not modified (up to taking their inverse). We illustrate this in a simple example with a single internal loop in section \ref{1loop}.

\smallskip

Since our definition of boundary spinors depend on the choice of tree that we use for gauge-fixing the region's bulk and it affects the closure defect, we should wonder about the meaning of this maximal tree. We have a few clear possibilities:

\begin{itemize}

\item The choice of tree is not physical since the external observer does  not have access to the region's bulk. Our definition of boundary spinors and closure defect are therefore not  physically-meaningful quantities and we should work more to identify true tree-independent coarse-grained observables. We don't know whether this is possible or not.

\item The choice of tree is not physical, therefore the boundary spinors and closure defects should not depend on the choice of tree when considering {\it physical states}. This would provide a well-defined mathematical criteria to define physical states in loop gravity. It should be related to the Hamiltonian constraints and the diffeomorphism invariance. Indeed the constraints that it would define, $\la z_{i}|G_{i}G_{j}^{-1}| z_{j}\ra=\la z_{i}|G_{i}\,\cG \,G_{j}^{-1}| z_{j}\ra$, probing the matrix elements of the holonomies on certain spinors, is exactly of the same type as the Hamiltonian constraints for 3d (loop) quantum gravity (and 4d BF theory) introduced in \cite{recursion3d,recursion4d, recursion_spinor}.

This underlines the relation between the diffeomorphism invariance and the coarse-graining of (quantum) states of geometry. For instance, in topological BF theory (which includes 3d quantum gravity), one can gauge out completely local fluctuations and project all the physical data onto the boundary then the coarse-graining of a region's bulk becomes trivial. In the context of quantum gravity, the holographic principle should play a similar role. 

\item The choice of tree is mathematically relevant and is linked to the external observer's definition of the region's boundary. For instance, it could be related to an ordering of the boundary edges and could somehow reflect an external tree structure. Indeed the observer needs to synchronize the reference frames at the vertices at the other end of the boundary edges in order to make meaningful measurements of the holonomies and curvature. This purpose requires choosing paths with trivial holonomies between the boundary edges, but in the external graph. The choice of internal might have to be related to the choice of an external tree, as in the analysis of entropy in topological BF theory \cite{danny-bf}, but this would require more thinking about the definition of an external observer.

\item The choice of tree is physically relevant. It corresponds to the paths in the bulk along which we probe the holonomy and thus the curvature. Imagining a thought experiment when an external observer sends some particles or matter beam through the bounded region and computes (some components of) the holonomy (or reference frame rotation) along the path followed by the beam (likely by measuring some interference pattern). It could follow any path within the region's bulk, it could even wind around some internal loops. The choice of path and thus on tree is then crucial. But it depends on how the beam or particle is sent to probe the region. Maybe the coarse-graining process should select the paths, and thus the tree, which is the most probable. Or maybe we should even keep track of all the possible trees (and thus internal structures) and consider a (quantum) statistical mixture of these as coarse-grained state (see for example the discussion in \cite{danny-cg}).

\item The choice of tree might be physically relevant or not, but this should merely be a criteria to identify what's a good region to coarse-grain. Indeed, we would to coarse-gain a region when we can neglect its internal structure. Therefore we should define a threshold for the dependence of the closure defect and boundary spinors on the tree $T$ beyond which we shouldn't coarse-grain the considered region. Ultimately, this would provide a threshold on (some components) of the curvature in the bulk: if the holonomies probed by the boundary edges are weak enough, then we can legitimately consider this region as a single vertex. 

\end{itemize}

Finally, the dependence of our boundary observables on the choice of internal tree is symptomatic of the existence of curvature and non-trivial holonomies within the bulk. Curvature is central to general relativity and we can not avoid this problem. The issue is, in broader words, to define a reference frame associated to the region's boundary. This does not seem obvious and the solution is probably a mixture of the scenarios considered above.

\section{The Boost Action}
\label{boost}

Let us consider a set of $N$ spinors $z_{i}$ around a single vertex, but possibly defined as the boundary spinors of a coarse-grained bounded region. We need to relax the closure constraints and consider  generic sets of spinors in $\C^{2N}$.
In that case we interestingly notice that the Casimir equations \eqref{casimirE} and \eqref{casimirF},
$$
\cE^{2}=\cA^{2}+|\vcC|^{2},
\qquad
\cF^{2}=\cA^{2}-|\vcC|^{2},
$$
giving the (related) quadratic  $\U(N)$ Casimirs $\cE^{2}$ and $\cF^{2}$ in terms of the boundary area $\cA$ and the closure defect, are very similar to the relativistic dispersion relation $E^{2}=m^{2}c^{4}+p^{2}c^{2}$. In fact the similarity is even more striking if we recall that the Casimir equations are derived from looking at the norm of Hermitian $2\times 2$ matrices which is simply the spinorial notation for 4-vectors. We will see below that the Casimir equations are indeed the dispersion relation for the $\SL(2,\C)$ action on the (boundary) spinors.

This naturally opens the door to many questions: what's the equivalent to the (Lorentz-invariant) rest mass $m$? to the energy $E$? to the speed of light $c$? Since the closure defect $|\vcC|$ is a measure of the curvature in the bulk, could it be effectively related to a quasi-local notion of mass or energy? Could this be applied to (rotating) black holes?  What's the space-time interpretation of the $\SL(2,\C)$ action on the spinors? We will not address these issues in detail here but focus on the mathematical expression of the boost action on the boundary spinors and how it allows to trivialize the closure defect. 

\subsection{Boosts, $\SL(2,\C)$-Invariants and Rest Area}

As introduced in \cite{un2} and investigated in detail in \cite{UN}, the global  action of $\SL(2,\C)$ transformations as $2\times 2$ matrices on the spinors is very interesting:
\be
 |z_{i}\ra \arr  |\tz_{i}\ra=G\, |z_{i}\ra,
 \qquad
G\in\SL(2,\C)\,.
\ee
A $\SU(2)$ group element is unitary, $G^\dagger=G^{-1}$, while a pure boost is Hermitian, $G^\dagger=G$.
The main mathematical statement is that the symplectic quotient $\C^{2N}//\SU(2)$ by the closure constraints is isomorphic to the quotient $\C^{2N}/\SL(2,\C)$ by the complexified action of $\SU(2)$ \cite{conrady,UN}. Moreover, the (holomorphic) observables $F_{ij}$ are $\SL(2,\C)$-invariant and label the $\SL(2,\C)$-orbits \cite{UN}.

This means that, starting with an arbitrary set of boundary spinors $z_{i}$, there exists a unique pure boost $\Lambda$ that maps it onto a set of spinors $\tz_{i}=\Lambda^{-1}\, z_{i}$ satisfying the closure constraints. This can be easily seen by directly considering the $2\times 2$ matrix $\cX=\sum_{i}  |z_{i}\ra\la z_{i}|$. This is a positive Hermitian matrix and we can define its square-root:
\be
\cX=\sqrt{\det \cX}\,\Lambda^{2},\qquad \Lambda=\Lambda^{\dagger}, \quad\det\Lambda =1\,.
\ee
Then we define the new set of spinors, $\tz_{i}=\Lambda^{-1}\, z_{i}$, which will automatically satisfy the closure constraints:
\be
\cX=\sum_{i}  |z_{i}\ra\la z_{i}|
\quad\longrightarrow\quad
\tX=\sum_{i}  |\tz_{i}\ra\la \tz_{i}|
=\Lambda^{-1}\cX\Lambda^{-1}{}^{\dagger}
=\sqrt{\det \cX} \,\id\,.
\ee
After a suitable $\SU(2)$ rotation to align the boost axis onto the $z$-direction, this pure boost appears simply as a global inverse rescaling of the two components of the spinors, which is enough to ensure the closure constraints are satisfied \cite{UN}.
Thus we can always map by a Lorentz transformation an arbitrary set of spinors, in particular boundary spinors defined by gauge-fixing of a region's bulk, to a new set of spinors satisfying the closure constraints. Moreover this map is unique (up to global $\SU(2)$ transformations). This allows to close any set of boundary sets without loss of information.

\smallskip

It is therefore interesting to check how the various observables change under boosts and identify the Lorentz-invariant observables. For instance, as insisted upon earlier, the holomorphic scalar products $F_{ij}$ are $\SL(2,\C)$-invariant\footnotemark:
\be
F_{ij}=[z_{i}|z_{j}\ra
\quad\longrightarrow\quad
\tF_{ij}=[\tz_{i}|\tz_{j}\ra
=[z_{i}|G^{-1}G|z_{j}\ra\,.
\ee
\footnotetext{
Le us remind that an arbitrary $2\times 2$ matrix $M$ always satisfies $M\eps M^{t}\eps^{-1}=(\det M)\id$.
Thus a Lorentz transformation $G$, with unit determinant, will satisfy:
$$
G^{-1}=\eps G^{t} \eps^{-1}=(\eps \bar{G} \eps^{-1})^{\dagger}\,.
$$
}
The determinant $\det X$ is also Lorentz-invariant:
$$
\cX\arr \tcX=G \cX G^{\dagger},
\qquad
\det \cX\arr \det\tcX= \det G \cX G^{\dagger}=\det \cX\,.
$$
This is not a new invariant since one checks the nice identity:
\be
\det \cX=\f12\tr \cX \eps \cX^{t}\eps^{-1}=\f12\tr \cX \eps \bcX\eps^{-1}
=\f12\sum_{ij} |F_{ij}|^{2}
=\cF^{2}\,.
\ee
This is important since $\det \cX=\det\tcX$ gives the area of the final closed spinors $\tz_{i}=\Lambda^{-1}z_{i}$:
$$
\widetilde{\cA}=\f12\tr\tcX=\sqrt{\det \tcX}=\cF\,.
$$ 
This $\SL(2,\C)$-invariant is exactly the equivalent of the rest mass, and we will thus refer to it as the {\it rest area}.

Reversely, starting from the closed spoors $\tz_{i}$ with the diagonal matrix $\tcX= \cF\,\id$ and boosting them to the initial spinor $z_{i}=\Lambda\,\tz_{i}$, with a pure boost parameterized as $\Lambda=\cosh \f\eta2\id+\sinh\f\eta 2 \hat{u}\cdot\vsigma$ with rapidity $\eta$ and arbitrary (irrelevant) axis $\hat{u}$, we get:
\beq
&&\cA=\f12\tr \cX=\f12 \tr\Lambda \tcX \Lambda^{\dagger}=\f12\cF\tr\Lambda^{2}=\cF\,\cosh \eta\,, \\
&&\cE^{2}=\f12\tr \cX^{2}=\cF^{2}\cosh^{2}\eta\,,\\
&& |\vcC|^{2}=\cE^{2}-\cA^{2}=\cF^{2}\sinh^{2}\eta\,,
\eeq
such that we recover a dispersion relation between the {\it rest area} $\cF$ (giving the boundary area of the closed spinors), the original  {\it boundary area} $\cA$ and the  {\it closure defect} $ |\vcC|$ measuring the curvature in the region's bulk:
\be
\cF^{2}=\cF^{2}\cosh^{2} \eta - \cF^{2}\sinh^{2} \eta
=\cA^{2}- |\vcC|^{2}\,.
\ee

\smallskip

The natural question in this context is whether the mathematically introduced rest area $\cF$ has a physical interpretation or not, or more generally these boosts acting on the (boundary) spinors have a space-time interpretation? For instance as boosts acting on some reference frame attached to the region or final coarse-grained vertex?
A possible scenario is that these boosts act on the time normal (to the canonical hypersurface) and change locally the 3d reference frame and the embedding of the spatial slice in space-time. Investigating further this possibility would require lifting spinor networks to twistor networks \cite{twistor1,twistor2}. Twistor networks  allow to explicitly represent the (discretized version of the) Lorentz space-time connection and the ``spinor$\hookrightarrow$ twistor'' embedding depends explicitly on the time normal field \cite{twistor3}.
At the continuum level, changing the hypersurface's embedding, and thus the extrinsic curvature, directly affects the curvature and holonomies of the Ashtekar-Barbero connection e.g. \cite{samuel}. Thus playing with the time normal field might allow to trivialize some holonomies in the bulk and define some boundary spinors satisfying the closure constraints (which would naturally follow from having only trivial holonomies around internal loops in the bulk). We postpone a detailed analysis to future investigation. 

\smallskip

Another attempt to a geometrical interpretation is provided by the null-vector approach developed in \cite{yasha}.
Considering the vectors $\vV_{i}\in\R^{3}$ defined from the spinors, we embed them in four-dimensional future-oriented null-vectors:
\be
v_{i}^{\mu}
\,\equiv\,
\la z_{i}|\sigma^{\mu}|z_{i}\ra
\,=\,
(V_{i}\,,\,\vV_{i})
\,=\,
(\la z_{i}|z_{i}\ra\,,\,\la z_{i}|\vsigma|z_{i}\ra)
\in\R^{4},
\qquad
v_{i}\cdot v_{i}=v_{i}^{\mu}v_{i\,\mu}=0\,,
\ee
or with the convention $\sigma^{0}=\id$. We do not assume that the vectors $\vV_{i}$ satisfy the closure constraints. We define instead the time-like direction given by the sum of all these null-vectors:
\be
n\equiv \sum_{i} v_{i},
\qquad
n^{\mu}=\f12 \tr \cX\sigma^{\mu} = (\cA,\vcC)\,.
\ee
This defines a common reference frame for all the 3-vectors. If we now project all the null vectors orthogonally to this time normal, the new vectors will automatically satisfy the closure constraints:
\be
W_{i}^{\mu}= v_{i}^{\mu}- \f{n\cdot v_{i}}{n\cdot n}n^{\mu}\,,
\qquad
\sum_{i}W_{i}=0\,.
\ee
Thus these new space-like vectors are all orthogonal to the time normal $n$ and live in the same spatial slice. They satisfy the closure constraints and define a  unique closed polyhedron. It is straightforward to check that this definition matches the previous one, that is $W_{i}^{\mu}=\la \tz_{i}|\sigma^{\mu}|\tz_{i}\ra$ in terms of the boosted spinors $\tz_{i}$ introduced above. This shows that if we consider a closed polyhedron and think of the normals to its faces as null-vectors, then there is a unique frame in which the polyhedron is actually closed. In all other frame, the sum of the 3d projections of the null-vectors do not close, we have a closure defect and we can not legitimately reconstruct the polyhedron.

This point of view might be particularly relevant to the study of black holes in lop gravity since the event horizon is indeed a null-surface. 

\smallskip

A detailed investigation of the geometrical meaning of the Lorentz action on the (boundary) spinors is crucial to understanding the coarse-grained of spin(or) networks in loop quantum gravity and specially for the study of quantum black holes (for example, our notion of $\SL(2,\C)$-invariant rest mass could be relevant to the definition of the horizon area, which is also Lorentz-invariant).

\subsection{The One-Loop Example}
\label{1loop}

\begin{figure}[h]
\includegraphics[height=25mm]{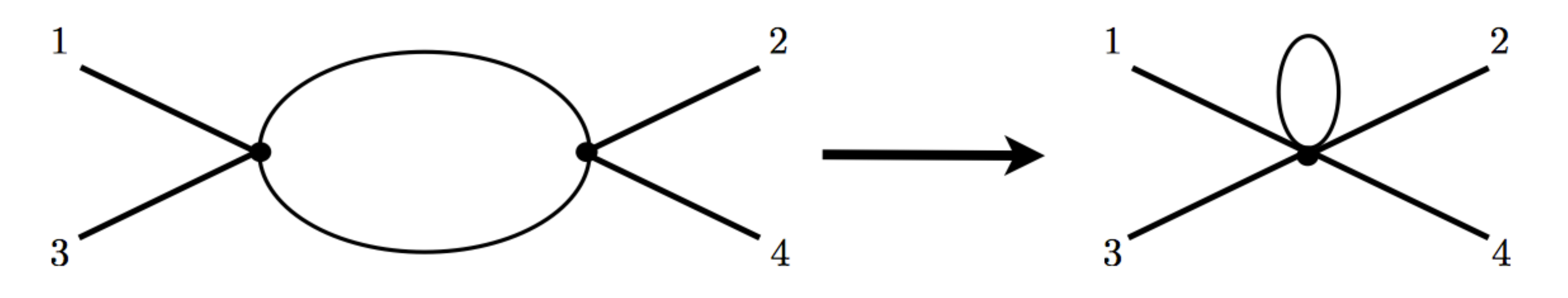}
\caption{\label{1loop4} The candy graph consists in $V=2$ internal vertices linked by $E=2$ internal edges, with $N=4$ boundary edges. It gets gauge-fixed to a four-valent vertex with one self-loop. 4-valent intertwiners are very interesting since they define dual tetrahedra. Here the internal loop will lead to curvature inside the dual tetrahedron and to an effective closure defect.}
\end{figure}

Let us illustrate our coarse-graining procedure with the simplest non-trivial internal graph - a single loop- with two vertices and two boundary edges. One could make it less simple and more ``realistic'' (allowing for a non-vanishing volume dual to the vertices) by considering two boundary edges attached to each vertex, as drawn on fig. fig.\ref{1loop4}, but this wouldn't change the present discussion on the gauge-fixing and coarse-graining procedure.

\begin{figure}[h]
\includegraphics[height=35mm]{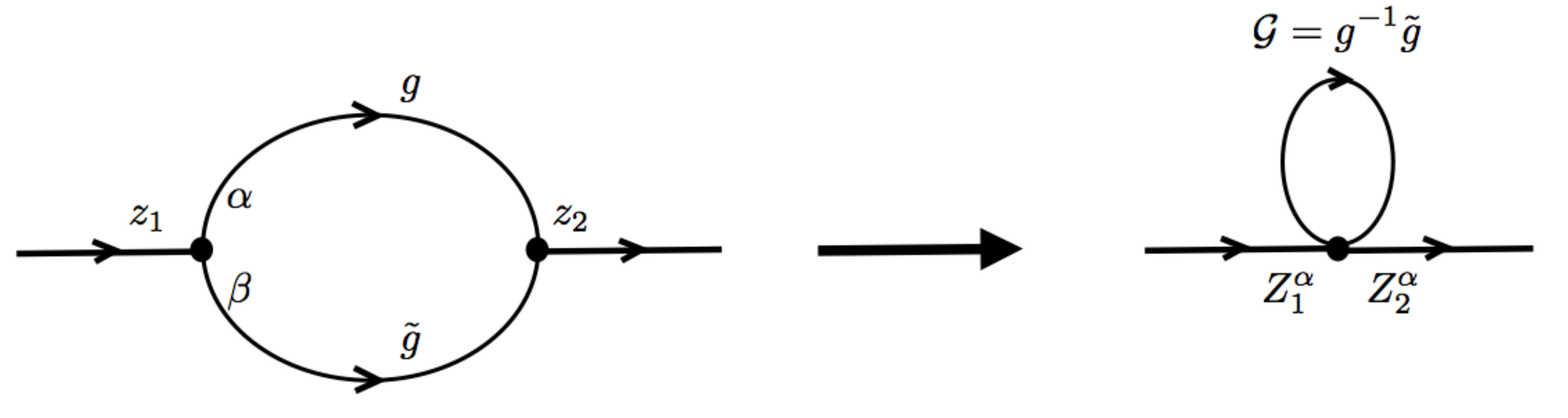}
\caption{\label{1loop2} One can simplify the candy graph down to the simplest case with only $N=2$ external edges. This corresponds to a single edge with one loop, similar to the one-loop correction to the propagator in standard quantum field theory. Here we gauge-fix the internal graph choosing the upper edge $\alpha$ as the tree. This leads to one self-loop carrying the holonomy $\cG=g^-1\tg$ and two boundary spinors $Z^\alpha_{1}$ and $Z^\alpha_{2}$.}
\end{figure}

As one can see on fig.\ref{1loop2}, one has two possible maximal trees for the single loop, either the upper link or the lower link. Choosing as reference vertex $v_{0}$, one defines the parallel-transported boundary spinors in both cases as:
\be
|Z_{1}^{\alpha}\ra=|z_{1}\ra, \quad
|Z_{2}^{\alpha}\ra=\,g^{-1}|z_{2}\ra, \qquad
\textrm{or}\qquad
|Z_{1}^{\beta}\ra=|z_{1}\ra, \quad
|Z_{2}^{\beta}\ra=\,\tg^{-1}|z_{2}\ra\,.
\ee
Focusing for now on the first choice of tree, one can compute the closure vector associated to the two boundary spinors. Using the associated vectors $\vV_{1,2}$ and keeping in mind that $g\vartriangleright \vV^{\alpha}=-\vW^{\alpha}$ and $\tg\vartriangleright \vec{V}^{\beta}=-\vec{W}^{\beta}$, we get:
\be
\vcC^{\alpha}=\vV_{1}^{\alpha}+\vV_{2}^{\alpha}=-(\id-g^{-1}\tg)\vartriangleright \vec{V}^{\beta}\quad\ne 0\,,
\ee
which a priori does not vanish when the holonomy $g^{-1}\tg$ around the internal loop is not trivial. We see that the closure defect directly  detects the presence of the internal loop and its value reflects the non-trivial holonomy in the bulk.
If we shift tree, we get:
\be
\vec{\cC}^{\beta}=\vV_{1}^{\beta}+\vV_{2}^{\beta}=-(\id-\tg^{-1}g)\vartriangleright \vec{V}^{\alpha}\,.
\ee
Its difference with the previous closure vector is:
\be
\vec{\cC}^{\beta}-\vcC^{\alpha}=
-(\id- \tg^{-1}g)\vartriangleright \vV_{2}^{\alpha}\,,
\ee
and grows with the holonomy $\tg^{-1}g$ and the boundary area.
%

The gauge-fixed graph is for both choices of a tree is tad-pole as drawn on fig.\ref{1loop2}. The holonomy around the loop is fixed up to taking to inverse, either $g^{-1}\tg $ or $\tg^{-1}g$. This non-trivial holonomy leads to the difference in the corresponding closure vectors. This difference is bounded  by  the total boundary area $\cA$ times the ($L^{2}$) norm $||\id-\tg^{-1}g||$ (while we recall that  the closure defect must itself be necessarily bounded by the boundary area).

Then a suitable (unique pure) boost will map the boundary spinors $Z_{1,2}$ (omitting the index $\alpha$) to aligned spinors (equal to each other's dual up to a phase) satisfying the closure constraints. In this special case with two spinors, we provide an explicit formula:
\be
\Lambda=\f{|Z_{1}\ra\la\Omega|+|Z_{2}\ra[\Omega|}{[Z_{1}|Z_{2}\ra}\quad\in\SL(2,\C),
\qquad
|\Omega\ra=\mat{c}{ 1 \\0},\quad |\Omega]=\mat{c}{ 0\\1 }.
\ee
This $\Lambda$ is a Lorentz transformation but not necessarily a pure boost. It maps our boundary spinors to the canonical basis on $\C^{2}$ which obvious satisfies the closure constraints:
$$
\cX\,\longrightarrow\,
\Lambda^{-1}\,\cX\,\Lambda^{-1}{}^{\dagger}
\,=\,
\Lambda^{-1}\,\left(|Z_{1}\ra\la Z_{1}|+|Z_{2}\ra\la Z_{2}|\right)\,\Lambda^{-1}{}^{\dagger}
\,=\,
\cF^{2}\,\left(|\Omega\ra\la \Omega|+|\Omega][\Omega|\right)
\,=\,
\cF^{2}\,\id\,,
$$
with $\cF^{2}=|\,[Z_{1}|Z_{2}\ra\,|^{2}=|[z_{1}|g^{-1}|z_{2}\ra|^{2}$ gives the rest area. By subtracting this to the boundary area $\cA^{2}$, we get the closure defect.

It is problematic that the choice of tree, $\alpha$ or $\beta$, affects our observables so much. This is a problem that will need to be addressed. In the meanwhile, to summarize, our region can be represented after gauge-fixing as a 4-valent vertex with two edges and a self-loop with depending on the choice of tree:
\be
\left|
\begin{array}{l}
|Z_{1}^{\alpha}\ra=|z_{1}\ra \\
|Z_{2}^{\alpha}\ra=\,g^{-1}|z_{2} \ra\\
G^{\alpha}= g^{-1}\tg
\end{array}
\right.
\qquad
\textrm{or}
\qquad
\left|
\begin{array}{l}
|Z_{1}^{\beta}\ra=|z_{1}\ra \\
|Z_{2}^{\beta}\ra=\,\tg^{-1}|z_{2} \ra\\
G^{\beta}= \tg^{-1} g
\end{array}
\right.
\ee
Then we can coarse-grain it by forgetting about the internal loop and keeping track of the induced closure defect either through the data of the original boundary area $\cA$ and the rest area $\cF$, or through the more complete data of $\cA$ and the Lorentz boost.

A worry about switching to the boosted spinors, even though it allows to keep satisfying the closure constraints, is that it will break the matching constraints along the boundary edges, since it affects the norm of the spinors. We might then have to consider generalized spinor networks without the matching constraints. This is actually a generic feature of the projected spin network basis used for spinfoam models \cite{projected1,projected2} to carry two $\SU(2)$ spins on each edge, living at both ends, and we will have to investigate how this could fit in the twistor network picture \cite{twistor3}.


\section{ On Coarse-Graining a Spin Network}

We now turn to the gauge-fixing and coarse-graining of the spin network wave-functions at the quantum level.
Starting with the graph $\Gamma$, we consider a generic gauge-invariant function of the holonomies along its edges, $\phi(\{g_{e}\}_{e\in\Gamma})$.
Considering a bounded region with internal graph $\Gamma_{in}$, we can gauge-fix this wave-function as described in the previous sections by choosing a reference vertex $v_{0}$ and a maximal tree $T$ on $\Gamma_{in}$. The gauge-invariance of the wave-functin $\phi$ ensures that we are not losing any information:
\be
\phi(\{g_{e}\}_{e\in\Gamma})
\,=\,
\phi(\{h_{t(e)}g_{e}(h_{s(e)})^{-1}\}_{e\in\Gamma})
\,=\,
\phi(\{\id\}_{e\in T},
\{\cG_{e}^{T}=G_{t(e)}^{-1}g_{e}G_{s(e)}\}_{e\in\Gamma_{in}\setminus T},
\{\tg_{e}\}_{e\in\pp\Gamma_{in}},\{g_{e}\}_{e\notin\Gamma_{in}})\,
\ee
where we have included the boundary edges in the definition of the internal graph $\Gamma_{in}$ an the gauge-fixed holonomies on boundary edges $\{\tg_{e}\}_{e\in\pp\Gamma_{in}}$ are either $G_{t(e)}^{-1}g_{e}$ or $g_{e}G_{s(e)}$ depending on whether their internal vertex is their source or target vertex.

As discussed in \cite{danny-cg}, the coarse-graining of the region would be to fix the holonomies around the internal loops $\cG_{e}^{T}$ to some specific values or more generally to integrate over them with some given probability distribution. More generally, this needs to be done with group averaging to the gauge-orbits $\{h\cG_{e}^{T}h^{-1}\}_{h\in\SU(2)}$ under the adjoint action of $\SU(2)$ in order to define a gauge-invariant coarse-grained wave-function. But then, how to determine the probability distribution for these internal holonomies?

We can simply integrate over the internal loops $\cG_{e}^{T}$ with the $\SU(2)$ Haar measure. However this would amount to project onto the component of the initial wave-function $\phi_{\Gamma}$ that has trivial spins $j_{e\in\Gamma_{in}}=0$ on the internal graph. This represents a trivial degenerate geometry within the region with no fluctuation of the connection. It is a brutal projection that kills any information about the bulk geometry. If one wants to probe the internal geometry, one should allow (slowly) for higher and higher spins and thus for fluctuations corresponding to higher modes  of the connection. But this corresponds merely to truncating the wave-functions with a spin cut-off. This could be a good way to bound the holonomy fluctuations within the region. It could be nevertheless interesting to determine instead what is the most probable holonomies in the bulk and evaluate the wave-function on these. In that case, it is the wave-function itself that dictates the amplitude probability. The best way to implement is to use the standard technique: {\it using the density matrix and tracing out over the bulk degrees of freedom}. This treats the bulk as an environment for the boundary degrees of freedom (and eventually the exterior state). In particular, at the dynamical level, we will then expect the boundary dynamics to be described by a master equation (of the Lindblad type) with probably a decoherence induced by tracing out the region's bulk. It wold be enlightening to understand which superselection sectors for spin networks this would lead to.

In this scenario, it is the bulk gravitational degrees of freedom and fluctuations that makes the system (or more exactly the description of the system by an external observer, i.e. the boundary degrees of freedom that an external observer has access to) decohere. Such a possibility should be investigated further and its application to quantum black holes would be interesting. 

\smallskip

Mathematically, the initial density matrix for a pure state is $\rho=| \phi\ra\la\phi |$ or more explicitly $\rho_{\Gamma}(g_{e},g'_{e})\,\equiv \overline{\phi(g_{e})}\,\phi(g'_{e})$. Coarse-graining by integrating over the group elements within the region (but not on the boundary edges) gives a mixed state for the holonomies outside (and on the boundary):
\be
\rho_{out}(\{g_{e},g'_{e}\}_{e\notin \Gamma_{in}})
\,\equiv\,
\int [dg_{e\in\Gamma_{in}}]\,\rho_{\Gamma}(\{g_{e},g'_{e}\}_{e\notin \Gamma_{in}}, \{g_{e},g_{e}\}_{e\in \Gamma_{in}})\,.
\ee
%
The subtle point of this definition is about the gauge-invariance properties of this reduced density matrix. Indeed, the initial state $\rho_{\Gamma}$ is invariant under independent $\SU(2)$ gauge transformations of the $g_{e}$ and $g'_{e}$ group elements. But due to the integration over the internal holonomies, the proposed reduced density matrix $\rho_{out}$ couples the $g_{e}$'s and $g'_{e}$'s and is invariant under only one set of $\SU(2)$ gauge transformations. Thus $\rho_{out}$  is not a statistical mixture of gauge-invariant spin network states. This symmetry breaking echoes the closure defect derived in the classical case.

In the case that we starts with a pure spin network states, labeled with spins $j_{e}$ on the edges and intertwiners $I_{v}$ at the vertices, if one integrate over the group elements $g_{e\in\Gamma_{in}}$ as prescribed above, one erases the internal region (internal vertices and edges) on both spin networks $\overline{\phi}(g_{e}$ and $\phi(g'_{e})$, living them with open boundary edges and then glues $\phi$ with $\overline{\phi}$ along these boundary edges with spins $j_{e\in\pp\Gamma_{in}}$, finally yielding a single doubled spin network state. This means that, either our definition is not consistent because not gauge-invariant, or we must allow for open spin network states, with open edges, which are not gauge invariant but covariant. This is the quantum counterpart of allowing for closure defect at the classical level. 

\smallskip

\begin{figure}[h]
\includegraphics[height=25mm]{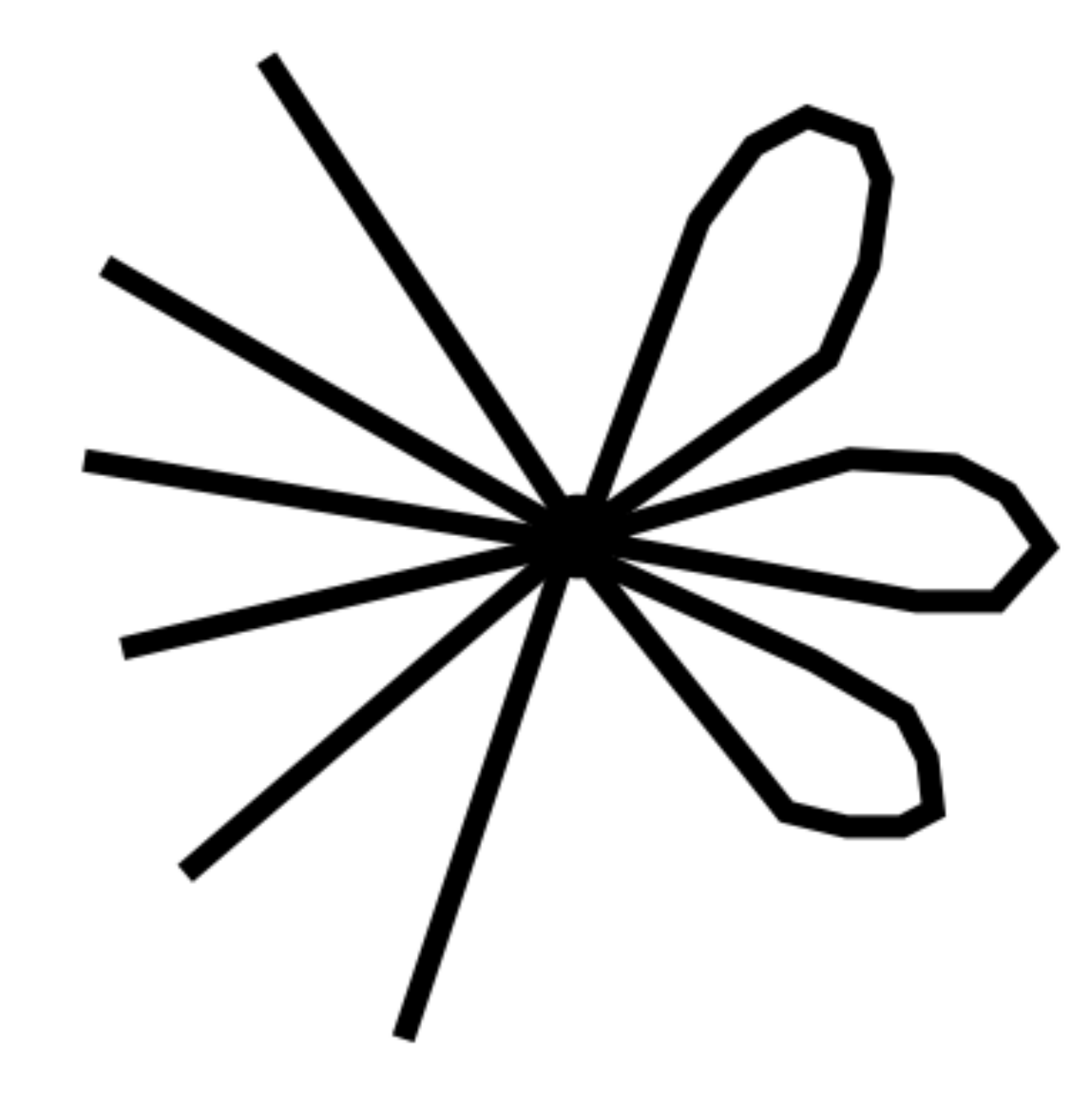}
\caption{\label{flower} The flower graph around a single vertex with many self-loops: the holonomies around the self-loops are noted $\cG_{p}$, the holonomies along the boundary edges (all oriented outward) are $g_{i}$ while we write $g_{e}$ for the holonomies along all the other edges.}
\end{figure}

One might argue that this issue arises because we are integrating over non-$\SU(2)$-invariant variables, hence brutally breaking the gauge invariance. In this case, one should first choose an internal tree $T$, gauge-fix the wave-function reducing the internal graph to a flower with petals, and integrate over the internal loop holonomies $\cG_{e}^{T}$. Let us forget for a moment the initial gauge-fixing phase and start directly with a flower graph, i.e a single internal vertex with many self-loops, as illustrated on fig.\ref{flower}.
%
%
Considering a gauge-invariant wave-function, $\phi(g_{i},\cG_{p},g_{e})=\phi(g_{i}h,h^{-1}\cG_{p}h,g_{e})$ for all $h\in\SU(2)$, with $\{g_{i}\}_{i=1..N}$ being the holonomies the boundary edges attached to the (single) internal vertex (chosen as their source vertex), $\{\cG_{p}\}_{p=1..P}$ being the holonomies around the loops attached to the vertex, and $\{g_{e}\}$ denoting all the holonomies living on the rest of the graph. The reduced density matrix would be defined by integration over the $\cG_{p}$ group elements:
\be
\rho_{red}(g_{i},g_{e},g'_{i},g'_{e})
\,\equiv\,
\int [d\cG_{p}] \,\overline{\phi}(g_{i},\cG_{p},g_{e})\,\phi(g'_{i},\cG_{p},g'_{e})\,.
\ee
As easily checked, this density matrix is invariant under the coupled gauge transformation $\rho_{red}(g_{i}h,g_{e},g'_{i}h,g'_{e})=\rho_{red}(g_{i},g_{e},g'_{i},g'_{e})$ for a single group element $h\in\SU(2)$ and is not invariant under decoupled gauge transformations $\rho_{red}(g_{i}h,g_{e},g'_{i}h',g'_{e})\ne\rho_{red}(g_{i},g_{e},g'_{i},g'_{e})$ with two independent group elements $h,h'\in\SU(2)$.
%
This is due to the fact that the boundary holonomies are transformed along with the internal loops by a $\SU(2)$ gauge transformation and that they can not be decoupled entirely from the bulk. 

To remedy this, one could group average by hand over the adjoint action on the internal loops:
$$
\rho_{red}^{int}(g_{i},g_{e},g'_{i},g'_{e})
\,\equiv\,
\int dh\,\int [d\cG_{p}] \,\overline{\phi}(g_{i},\cG_{p},g_{e})\,\phi(g'_{i},h^{-1}\cG_{p}h,g'_{e})\,.
$$
This extra group averaging allows to decouple $\phi$ from $\bar{\phi}$ and ensures the invariance under decoupled gauge transformations, as expected for statistical mixture of spin network states. However, such an extra group averaging does not have a clear physical motivation or interpretation.

\smallskip

To summarize, the same issue arises at the quantum level than in the classical setting: one has to provide a physical interpretation to the closure defect (mass? quasi-local energy?) or a clear physical reason to project it out.

One should also study these questions from the more covariant viewpoint of projected spin networks used in spinfoam models \cite{projected1,projected2}, which allow for boosting the time normal and explicitly 
playing with the hyper surface embedding. 

%
%
%
%


\section{Conclusion \& Outlook}

In the present work, we have investigated the boundary observables and deformations of a bounded region of a spinor network in loop gravity, leading us to discuss the coarse-graining of such a region to a single vertex. By a gauge-fixing of the local $\SU(2)$ transformations, we were able to effectively contract an arbitrary region to a single vertex, with its boundary edges plus some self-loops coming from all the (independent) loops of the original internal graph of the region.The difference with a single vertex is exactly these self-loops, reflecting the internal structure and degrees of freedom of the region's bulk.  This allowed to identify the $\SU(2)$-invariant boundary observables and to show that their Poisson-algebra is exactly the same as in the case of a single vertex. In particular, the area-preserving deformations are once again identified as $\U(N)$ transformations, where $N$ is the number of boundary edges. The coarse-graining of the region then amounts to neglecting the self-loops or integrating over them. However, their existence and the fact that they can carry non-trivial holonomies implies that the boundary spinors do not satisfy the closure constraints anymore. In the case of a single vertex, the closure constraints ensure the existence of a unique dual polyhedron (with $N$ faces) embedded in flat 3d space. When the closure constraints are relaxed, this is not possible anymore (at least the straightforward correspondence is not). This does not mean that the region's boundary is not a closed surface anymore but reflects that the region's bulk has a non-trivial curvature. We further show that the value of the closure defect is related to the boundary area and the Casimir of the $\u(N)$ generators (identified as boundary diffeomorphisms at the discrete level).

All this carries through to the quantum level. Spinor networks get quantized to the spin network wave-functions of loop quantum gravity. Spinors satisfying the closure constraints around a single vertex become $\SU(2)$ intertwiners. And the space of intertwiners at fixed total boundary area carries an irreducible representation of the unitary group $\U(N)$. Coarse-graining a region of a spin network again leads to relaxing the closure constraints around the effective single vertex. The resulting state intertwines between the $\SU(2)$ representations living on the boundary edges and an extra one representing the closure defect. And we obtain $\U(N)$ irreducible representations labeled by the values of the total boundary area and the closure defect.

Finally, we show that there are natural Lorentz transformations with $\SL(2,\C)$ acting on the (boundary) spinors. The four component vector $(\cA,\vcC)$ with the boundary area and the closure 3-vector transform as a standard 4-vector  under these $\SO(3,1)$ Lorentz boosts. In particular, there exists a unique pure boost that maps the boundary spinors onto a configuration satisfying the closure constraints and thus defining unique dual convex polyhedron. However, these boosts change the boundary area (and in fact all the individual areas of the faces). Nevertheless we introduce the rest area $\cF=\sqrt{\cA^{2}-|\vcC|^{2}}$, which is a Lorentz-invariant and also the boundary area of the boosted closed spinor configuration. It is moreover a $\U(N)$-Casimir. We believe that this concept of Lorentz-invariant rest area could be relevant to the study of quantum black holes in the context of loop gravity.

\smallskip

There are three main issues raised by the present analysis:
\begin{itemize}
\item The physical interpretation of the closure defect:

How is the closure defect explicitly related to the bulk curvature? Can it be interpreted a local notion of mass or energy in loop gravity? This should be investigated in the continuum theory, for example for the Schwarzschild metric, taking into account that the vectors are a discretization of the triad field.

\item The physical interpretation of the Lorentz transformations:

Is there a space-time interpretation to the Lorentz boosts acting on the spinor variables and thus on the area and closure defect? Mostly likely, we need to analyze this from the point of view of twistor networks (and projected spin networks at the quantum level) and it looks as if they will act on the embedding of the canonical hypersurface within the space-time.

\item The ambiguity in the gauge-fixing procedure:

All the definitions of boundary observables and deformations depend on the choice of a maximal tree $T$ within the internal graph of the region. Should we require tree-independence, by either looking for better observables or by using it as a criteria to identify good physical states? Or should we look for a physical interpretation for the choice of tree, for example in terms of measurements made by an external observer to probe the region's geometry? Overall, it seems that tree-(in)dependence is the diffeomorphism independence in loop gravity (or a implementation of it at the discrete level). From this perspective, we would like to propose the definition of a new equivalence relation at both classical and quantum level: on a flower graph with boundary edges and self-loops, two configurations (spinors up to $\SU(2)$ gauge transformations or intertwiners) are equivalent off there exist more refined graph and state from which can be derived these two configurations by different choices of tree. This means that the two configurations represent coarse-grainings of a same microscopic state. Natural questions are whether this is linked to diffeomorphisms or not? whether we should require the loop gravity dynamics to be invariant under such equivalence relation? 

\end{itemize}

The dependence of the gauge-fixing procedure, and thus of our coarse-grained observables, on the choice of the tree $T$ in the internal graph is a crucial issue. Although it seems at first an entirely mathematical matter, it easily acquires a physical dimension. It defines a set of basic paths between the vertices of the region and there is indeed a unique path between every two boundary vertices along a given tree. Imagining a particle or wave going through the considered region, it would be best suited to choose its path within the region as part of the tree and synchronize the reference frames along its trajectory. However, if we are coarse-graining the region in that setting, it would be because the particle's size is larger than the region's length scale, in which case this becomes similar to the two-slit experiment of quantum mechanics. Following this similarity, it would seem more reasonable to assume that the particle's wave-function explores all the possible paths within the region, in which case it seems better to consider the reduced density matrix summing over all possible choices of internal tree $T$ and hope for some decoherence phenomenon of the ``particle+region'' in a semi-classical regime.

\smallskip

From here, there are a few things that could be investigated. First, we could study how to refine the gauge-fixing  and the resulting coarse-graining procedure. We could work with a double tree-structure (representing the inside region and the outside space) as in \cite{danny-bf} or directly with a mutli-tree structure implementing the coarse-graining the whole spatial slice as once. We can also re-consider the space of spinor networks and spin network wave-functions as the the set of configurations for loop gravity. It is impossible to avoid the existence of the closure defect. On the one hand, we can try to re-close the spin(or) network. Either by a brute-force group averaging thus setting the closure defect to 0 and erasing all the internal loops by hand, or by using the Lorentz boosts. The first possibility is not (yet) well-motivated and erases the internal structure and curvature by hand, while the latter alternative mean upgrading the spin(or) networks to structures covariant under $\SL(2,\C)$ such as the twistor networks. On the other hand, we could decide to enrich the spin(or) networks with further data and we would like to propose two possible extensions of spin networks:
\begin{itemize}
\item Tagged spin networks:

We can generalize spin networks by relaxing the $\SU(2)$ gauge invariance at every vertex. More precisely, we add an extra open link to every vertex, a {\it tag}, to account for the closure defect. We would retain the $\SU(2)$ gauge invariance with $\SU(2)$ transformations acting on both the graph edges and the tags.

\item Loopy spin networks: 

Since self-loops are the gauge-fixed counterpart of the internal loops, we imagine a different structure for loop quantum gravity. Instead of defining a loop gravity dynamics that induce transitions between states living on different graphs, one could start with a fixed background graph (reflecting a certain sampling of space, possibly as probed and measured by a given observer) with dynamics acting on that fixed graph but possibly creating and annihilating self-loops at every vertex of that graph. We could then work on a Fock space over the background graph, with spin network states living on that background graph but also on any extension of it with extra self-loops. These self-loops accounts for the (possible) internal structure of each vertex and represent the coarse-graining of any more refined graph that could be obtained by the dynamics from the background graph.

\end{itemize}

These two generalizations of the  Hilbert space of spin networks will investigated in a separate work \cite{inprep}.

\smallskip

Our gauge-fixing and coarse-graining procedure should be relevant to studying the properties and dynamics of quantum black holes in loop gravity. Indeed black holes come from highly dense and curved regions of space(-time), but admit a naturally coarse-grained description since the external observer a priori ignores everything about the internal geometry of the region within the horizon. Following our gauge-fixing procedure, it is straightforward to reduce any internal region to a single vertex and one could imagine the black hole regime as states with very large number of self-loops (and possibly maximal closure defect). Then on might ask whether the loop gravity dynamics decouples the internal and boundary dynamics in this limit of infinite number of self-loops. In this context, we would like to push forward a possible link between our description of area-preserving deformations as $\U(N)$ transformations and the Carlip's picture for black hole entropy from boundary-preserving diffeomorphisms form in a Virasoro algebra \cite{carlip}. Indeed, in the limit $N\arr\infty$ (interpreted as an infinite refinement of the boundary or black hole horizon), we can identify a natural Witt sub-algebra of $\u(\infty)$. More precisely, for edge labels taking any integer value, we have:
\be
[E_{ij},E_{kl}] =(\delta_{jk}E_{il}-\delta_{il}E_{kj})
\quad\Longrightarrow\qquad
L_{n}\,\equiv\,\sum_{k}k E_{k,n+k},\quad
[L_{n},L_{m}]=(n-m)L_{n+m}\,.
\ee
In fact, the $\u(N)$ generators $E_{ij}$ are constructed from the spinor variables and we can define two Witt algebra, one for each of the two spinor components.
Natural questions are whether the Virasoro central extension of this sub-algebra makes sense for the $\u(N)$ algebra from our gauge-fixed point of view  and whether this has anything to do with Carlip's approach. 

\smallskip

More general, one can wonder what happens when coarse-graining a superposition of graph, or defining boundary deformations for a coherent mesh of graphs, like arising in the recent work on black holes \cite{alej-recent} modeling the horizon with superpositions of different $N$'s. To this purpose, we need to investigate the consistency between our gauge-fixing procedure and the projective limit tool used to define the Hilbert space of loop quantum gravity \cite{projective}.

Finally, we would like to conclude with a last thought.
The coarse-graining should slowly change the background, or more exactly update the initially trivial background with the spin network state data, thus separating the  geometrical background (large scale structure) from the field fluctuations (small scale structure) during the coarse-graining process. This seems to require maps between the Hilbert spaces for loop quantum gravity with a non-trivial vacuum \cite{hanno}. These states might turn out to be  interpretable in terms of tagged or loopy spin networks.

\section*{Acknowledgements}

I would thank Simone Speziale and Carlo Rovelli for some very motivating discussions on the physical relevance of the gauge-fixing procedure and on the coarse-graining of spin networks.




\begin{thebibliography}{99}

\bibitem{lqg-review1}
M. Gaul and C. Rovelli,
{\it Loop Quantum Gravity and the Meaning of Diffeomorphism Invariance},
Lect.Notes Phys. 541 (2000) 277-324 [arXiv:gr-qc/9910079]

\bibitem{lqg-review2}
C. Rovelli,
{\it Quantum Gravity},
Cambridge Monographs on Mathematical Physics (2007), Cambridge University Press

\bibitem{lqg-review3}
T. Thiemann,
{\it Modern Canonical Quantum General Relativity},
Cambridge Monographs on Mathematical Physics (2008), Cambridge University Press

\bibitem{twisted1}
L. Freidel and S. Speziale,
{\it Twisted geometries: A geometric parametrisation of SU(2) phase space},
Phys.Rev.D82 (2010) 084040 [arXiv:1001.2748]

\bibitem{spinor}
E.F. Borja, L. Freidel, I. Garay and E.R. Livine,
{\it U(N) tools for Loop Quantum Gravity: The Return of the Spinor},
Class.Quant.Grav.28 (2011) 055005 [arXiv:1010.5451]

\bibitem{spinor_johannes}
E.R. Livine and J. Tambornino,
{\it Spinor Representation for Loop Quantum Gravity},
J. Math. Phys. 53, 012503 (2012) [arXiv:1105.3385]

\bibitem{spinor_holo}
E.R. Livine and J. Tambornino,
{\it Holonomy Operator and Quantization Ambiguities on Spinor Space},
Physical Review D87 (2013) 104014  [arXiv:1302.7142]

\bibitem{UN}
E.R. Livine,
{\it Deformations of Polyhedra and Polygons by the Unitary Group},
arXiv:1307.2719

\bibitem{un0}
F. Girelli and E.R. Livine,
{\it Reconstructing Quantum Geometry from Quantum Information: Spin Networks as Harmonic Oscillators},
Class.Quant.Grav. 22 (2005) 3295-3314 [arXiv:gr-qc/0501075]

\bibitem{un1}
L. Freidel and E.R. Livine,
{\it The Fine Structure of SU(2) Intertwiners from U(N) Representations},
J.Math.Phys. 51 (2010) 082502 [arXiv:0911.3553]

\bibitem{un2}
L. Freidel and E.R. Livine,
{\it U(N) Coherent States for Loop Quantum Gravity},
J.Math.Phys.52 (2011) 052502 [arXiv:1005.2090]

\bibitem{un3}
M. Dupuis and E.R. Livine,
{\it Revisiting the Simplicity Constraints and Coherent Intertwiners},
Class.Quant.Grav. 28 (2011) 085001 [arXiv:1006.5666]

\bibitem{bhentropy_danny}
E.R. Livine and D. Terno,
{\it Entropy in the Classical and Quantum Polymer Black Hole Models},
arXiv:1205.5733

\bibitem{bh_karim1}
J. Engle, K. Noui, A. Perez and D. Pranzetti,
{\it Black hole entropy from an SU(2)-invariant formulation of Type I isolated horizons},
Phys. Rev. D82 (2010) 044050 [arXiv:1006.0634]

\bibitem{bh_karim2}
J. Engle, K. Noui, A. Perez and D. Pranzetti,
{\it The SU(2) Black Hole entropy revisited},
arXiv:1103.2723

\bibitem{bh_kaul_review}
R. Kaul,
{\it Entropy of Quantum Black Holes},
SIGMA 8 (2012) 005 [arXiv:1201.6102]

\bibitem{bh_jacobo_review}
J. Diaz-Polo and D. Pranzetti,
{\it Isolated Horizons and Black Hole Entropy In Loop Quantum Gravity},
arXiv:1201.6102

\bibitem{un4}
M. Dupuis and E.R. Livine,
{\it Holomorphic Simplicity Constraints for 4d Spinfoam Models},
Class.Quant.Grav. 28 (2011) 215022 [arXiv:1104.3683]

\bibitem{un4_conf}
M. Dupuis and E.R. Livine,
{\it Holomorphic Simplicity Constraints for 4d Riemannian Spinfoam Models},
Conference Proceedings of Loops '11 (Madrid, Spain, 2011), to appear in Journal of Physics: Conference Series (JPCS) [arXiv:1111.1125]

\bibitem{2vertex}
E.F. Borja, J. Diaz-Polo, I.Garay and E.R. Livine,
{\it Dynamics for a 2-vertex Quantum Gravity Model},
Class.Quant.Grav.27 (2010) 235010 [arXiv:1006.2451]

\bibitem{2vertex_conf}
E.F. Borja, J. Diaz-Polo, L. Freidel, I.Garay and E.R. Livine,
{\it Dynamics for a simple graph using the U(N) framework for loop quantum gravity},
Conference Proceedings of Loops'11 (Madrid, Spain, 2011), to appear in Journal of Physics: Conference Series (JPCS), [arXiv:1110.6017]

\bibitem{recursion}
V. Bonzom, E.R Livine and S. Speziale,
{\it Recurrence relations for spin foam vertices},
Classical and Quantum Gravity 27 (2010) 125002

\bibitem{recursion3d}
V. Bonzom and L. Freidel,
{\it The Hamiltonian constraint in 3d Riemannian loop quantum gravity},
Class.Quant.Grav.28 (2011) 195006 [arXiv:1101.3524]

\bibitem{recursion4d}
V. Bonzom,
{\it Spin foam models and the Wheeler-DeWitt equation for the quantum 4-simplex},
Phys.Rev.D84 (2011) 024009 [arXiv:1101.1615]

\bibitem{recursion_spinor}
V. Bonzom and E.R. Livine,
{\it A new Hamiltonian for the Topological BF phase with spinor networks},
arXiv:1110.3272

\bibitem{SFcosmo_merce}
E.R. Livine and M. Mart\'in-Benito,
{\it Classical Setting and Effective Dynamics for Spinfoam Cosmology},
arXiv:1111.2867

\bibitem{spinor_conf}
E.R. Livine and J. Tambornino,
{\it Loop gravity in terms of spinors},
Conference Proceedings of Loops '11 (Madrid, Spain, 2011), to appear in Journal of Physics: Conference Series (JPCS), [arXiv:1109.3572]

\bibitem{twisted2}
L. Freidel and S. Speziale,
{\it From twistors to twisted geometries},
Phys.Rev.D82 (2010) 084041 [arXiv:1006.0199]

\bibitem{noncompact}
L. Freidel and E.R. Livine,
{\it Spin Networks for Non-Compact Groups},
J.Math.Phys. 44 (2003) 1322-1356 [arXiv:hep-th/0205268]

\bibitem{polyhedron}
E. Bianchi, P. Dona and S. Speziale,
{\it Polyhedra in loop quantum gravity},
Phys.Rev.D83 (2011) 044035 [arXiv:1009.3402]

\bibitem{twisted_cpt}
H.M. Haggard, C. Rovelli, F. Vidotto and W. Wieland,
{\it The spin connection of twisted geometry},
 arXiv:1211.2166
 
\bibitem{marc}
L. Freidel, M. Geiller and J. Ziprick,
{\it Continuous formulation of the Loop Quantum Gravity phase space},
 Class. Quantum Grav. 30 (2013) 085013 [arXiv:1110.4833]

\bibitem{spinning}
L. Freidel and J. Ziprick,
{\it Spinning geometry = Twisted geometry},
arXiv:1308.0040

\bibitem{generating}
V. Bonzom and E.R. Livine,
{\it Generating Functions for Coherent Intertwiners},
arXiv:1205.5677

\bibitem{danny-cg}
E.R. Livine and D. Terno,
{\it Reconstructing Quantum Geometry from Quantum Information: Area Renormalisation, Coarse-Graining and Entanglement on Spin Networks},
 arXiv:gr-qc/0603008
 
\bibitem{danny-bulk}
E.R. Livine and D. Terno,
{\it Bulk Entropy in Loop Quantum Gravity},
Nucl.Phys.B794 (2008) 138-153 [arXiv:0706.0985]

\bibitem{danny-bf}
E.R. Livine and D. Terno,
{\it The entropic boundary law in BF theory},
Nucl.Phys.B806 (2009) 715-734 [arXiv:0805.2536]

\bibitem{conrady}
F. Conrady and L. Freidel,
{\it Quantum geometry from phase space reduction},
 J.Math.Phys.50 (2009) 123510 [arXiv:0902.0351]
 
 \bibitem{twistor1}
M. Dupuis, L. Freidel, E.R. Livine and S. Speziale,
{\it Holomorphic Lorentzian Simplicity Constraints},
J. Math. Phys. 53 (2012)  032502 [arXiv:1107.5274]

\bibitem{twistor2}
E.R. Livine, S. Speziale and J. Tambornino,
{\it Twistor Networks and Covariant Twisted Geometries},
Phys. Rev. D 85 (2012)  064002  [ arXiv:1108.0369]

\bibitem{twistor3}
S. Speziale and W. Wieland,
{\it The twistorial structure of loop-gravity transition amplitudes},
arXiv:1207.6348

\bibitem{samuel}
J. Samuel,
{\it Is Barbero's Hamiltonian formulation a Gauge Theory of Lorentzian Gravity?},
Class.Quant.Grav.17 (2000) L141-L148 [arXiv:gr-qc/0005095]

\bibitem{yasha}
Y. Neiman,
{\it Polyhedra in spacetime from null vectors},
arXiv:1308.1982

\bibitem{projected1}
E.R. Livine,
{\it Projected Spin Networks for Lorentz connection: Linking Spin Foams and Loop Gravity},
Class.Quant.Grav. 19 (2002) 5525-5542 [arXiv:gr-qc/0207084]

\bibitem{projected2}
M. Dupuis and E.R. Livine,
{\it Lifting SU(2) Spin Networks to Projected Spin Networks},
 Phys. Rev. D 82 (2010) 064044 [arXiv:1008.4093]

\bibitem{inprep}
E.R. Livine,
{\it Tagged and Loopy Spin Networks for Loop Gravity},
in preparation

\bibitem{carlip}
S. Carlip,
{\it Effective Conformal Descriptions of Black Hole Entropy},
Entropy 2011, 13(7), 1355-1379 [arXiv:1107.2678]

\bibitem{alej-recent}
A. Ghosh, K. Noui and A. Perez,
{\it Statistics, holography, and black hole entropy in loop quantum gravity},
arXiv:1309.4563

\bibitem{projective}
A. Ashtekar and J. Lewandowski,
{\it Projective Techniques and Functional Integration},
J.Math.Phys.36 (1995) 2170-2191 [arXiv:gr-qc/9411046]

\bibitem{hanno}
T. Koslowski and H. Sahlmann,
{\it Loop Quantum Gravity Vacuum with Nondegenerate Geometry},
SIGMA 8 (2012), 026 [arXiv:1109.4688]




%
%
%
%
%
%
%



\end{thebibliography}
\end{document}